\documentclass{emulateapj}

\shortcites{schuberth06,jones02,RC3,humphrey06,zhang07,fukazawa06,kim06,paturel03,sivakoff07,woodley08,jordan04,sarazin03,douglas07,woodley07,dekel05,romanowsky03,churazov08,cote03,bergond06,gebhardt00,ghez98,genzel00,jarrett03,lokas07,finoguenov06,jetha07,sazonov06,brough06b,temi03,trinchieri94,sanchez-blazquez06,romanowsky08,bell03,spitler08a,posson08}

\newcommand{\ASCA}{\textit{ASCA}}
\newcommand{\caldb}{\textsc{caldb}}
\newcommand{\Chandra}{\emph{Chandra}}
\newcommand{\chisq}{\ensuremath{\chi^2}}
\newcommand{\ciao}{\textsc{ciao}}
\newcommand{\erg}{\ensuremath{\mbox{~erg}}}
\newcommand{\ergps}{\ensuremath{\erg \ps}}
\newcommand{\keV}{\ensuremath{\mbox{~keV}}}
\newcommand{\kpc}{\ensuremath{\mbox{~kpc}}}
\newcommand{\NH}{\ensuremath{N_{\mathrm{H}}}}
\newcommand{\ps}{\ensuremath{\s^{-1}}}
\newcommand{\Rosat}{\emph{Rosat}}
\newcommand{\rproj}{\textsc{r project}}
\newcommand{\s}{\ensuremath{\mbox{~s}}}
\newcommand{\sspline}{\textsc{smooth.spline}}
\newcommand{\XMM}{\emph{XMM-Newton}}
\newcommand{\XSPEC}{\textsc{xspec}}

\shorttitle{Comparing mass profiles in NGC 4636}
\shortauthors{Johnson et al.}
\begin{document}

\title{Comparing X-ray and dynamical mass profiles in the early-type galaxy NGC~4636}

\author{Ria Johnson\altaffilmark{1}, Dalia Chakrabarty\altaffilmark{2}, Ewan O'Sullivan\altaffilmark{3} and Somak Raychaudhury\altaffilmark{1}}
\email{ria@star.sr.bham.ac.uk}
\altaffiltext{1}{School of Physics and Astronomy, University of Birmingham, Edgbaston, Birmingham B15 2TT, UK}
\altaffiltext{2}{School of Physics and Astronomy, University of Nottingham, Nottingham NG7 2RD, UK\\ Current e-mail: D.Chakrabarty@warwick.ac.uk}
\altaffiltext{3}{Harvard-Smithsonian Center for Astrophysics, 60 Garden Street, Cambridge, MA 02138, USA}

\begin{abstract}
We present the results of an X-ray mass analysis of the early-type galaxy NGC~4636, using \Chandra\ data. We have compared the X-ray mass density profile with that derived from a dynamical analysis of the system's globular clusters (GCs). Given the observed interaction between the central active galactic nucleus and the X-ray emitting gas in NGC~4636, we would expect to see a discrepancy in the masses recovered by the two methods. Such a discrepancy exists within the central $\sim$10\,kpc, which we interpret as the result of non-thermal pressure support or a local inflow. However, over the radial range $\sim$\,10--30\,kpc, the mass profiles agree within the 1\,$\sigma$ errors, indicating that even in this highly disturbed system, agreement can be sought at an acceptable level of significance over intermediate radii, with both methods also indicating the need for a dark matter halo. However, at radii larger than 30\,kpc, the X-ray mass exceeds the dynamical mass, by a factor of 4--5 at the largest disagreement. A Fully Bayesian Significance Test finds no statistical reason to reject our assumption of velocity isotropy, and an analysis of X-ray mass profiles in different directions from the galaxy centre suggests that local disturbances at large radius are not the cause of the discrepancy. We instead attribute the discrepancy to the paucity of GC kinematics at large radius, coupled with not knowing the overall state of the gas at the radius where we are reaching the group regime ($>$30\,\kpc), or a combination of the two. 
\end{abstract}
\section{Introduction}
\label{sec:intro}
The current paradigm of galaxy formation describes how galaxies form embedded in massive dark matter halos. Whereas the measurement of rotation curves can be successfully applied to late-type galaxies to infer the presence of this dark matter \citep[see e.g.][for a review]{sofue01}, this cannot be employed in early-type galaxies as their stars and gas are not supported by rotation. Therefore, different methods must be invoked to measure the galaxy mass.

It has long been known that early-type galaxies contain hot ($\sim$10$^{6}$K) X-ray emitting gas \citep{forman85}, the temperature and density of which allow the determination of the total gravitating mass, assuming hydrostatic equilibrium and spherical symmetry. This approach has proved successful at yielding meaningful mass profiles for early-type galaxies \citep[e.g.][]{osullivan04b,fukazawa06,humphrey06,zhang07}. The effect of the assumption of spherical symmetry has been addressed in the case of galaxy clusters, indicating that although compression and elongation along the line-of-sight can under or over-estimate the central mass respectively, this is only a small effect at large radius \citep{piffaretti03}. However, the validity of the intrinsic assumption of hydrostatic equilibrium has been questioned with specific reference to early-type galaxies \citep{diehl06}. NGC~4636 presents an ideal test-bed in this respect, as it is a highly disturbed system, with evidence of bubbles and shocks caused by previous AGN outbursts \citep{jones02,ohto03,osullivan05b}.

The use of dynamical tracers of the gravitational potential is also a well-established method to recover the kinematics of bound systems, both on the scale of globular clusters \citep[e.g.][in M15]{gebhardt00}, and for the Galaxy itself \citep{chakrabarty01,genzel00,ghez98}. The use of globular clusters (GCs) as tracers of the potential has been particularly successful in recovering mass profiles of nearby elliptical galaxies \citep[examples include][]{romanowsky01,cote03,bergond06,schuberth06,woodley07}. Similarly, dedicated surveys of planetary nebulae in early-type galaxies can also be used to derive the distribution of matter \citep{douglas07}, although care is required to avoid complications from distinct populations of planetary nebulae, which have been seen for example in the galaxy NGC 4697 \citep{sambhus06}. These approaches involve solving the Jeans equations under the assumption of spherical symmetry to determine the galaxy mass. Interestingly, a study of three early-type galaxies using planetary nebulae kinematics, \citet{romanowsky03} concluded a significant lack of dark matter in these systems. However, \citet{dekel05} showed these data to be consistent with a massive dark halo when more radial orbits were considered. This highlights the mass-anisotropy degeneracy present in this approach, which can be broken by considering higher order velocity moments \citep{lokas07}. A further systematic effect is the assumption of spherical symmetry. In the case where the galaxy is flattened along the line-of-sight, its mass can be under-estimated if the system is assumed to be spherically symmetric \citep{magorrian01}.

As both X-ray and dynamical methods have their own intrinsic assumptions, the most robust constraints can be placed on the mass profiles of early-type galaxies when different methods are compared. Indeed, there is currently an emerging attempt to use different techniques in a complementary manner \citep[][]{romanowsky08,churazov08,samurovic06,bridges06}; additionally, this approach improves our understanding of the systematics involved in each method. Recent work by \citet{churazov08} explored in detail the comparison between X-ray and optically derived profiles for M87 and NGC 1399, finding agreement between the methods at the 10--20\,\% level when looking at the gravitational potential. However, both of these systems reside at the centres of clusters, M87 being the centre of Virgo, and NGC 1399 the centre of Fornax, and in both cases, the measurement of the potential is probing the cluster potential. In so-called `normal' elliptical galaxies, the situation is much less certain. For example, in the galaxy NGC 3379, \citet{pellegrini06} require an outflow of the X-ray emitting gas to bring the X-ray results into agreement with the optically derived results. Only by the study of more systems with multiple approaches will we be able to reconcile the observed discrepancies. This is successful on a local scale, as individual GCs and/or planetary nebulae need to be resolved, limiting the distance to which these observations can be made. Investigating the wider properties of the dark matter halos of elliptical galaxies will require techniques such as stacked lensing \citep[e.g.][]{sheldon04,hoekstra05,kleinheinrich06,koopmans06,mandelbaum06,ferreras08}.

The layout of the paper is as follows. Section \ref{sec:N4636} describes the basic properties of NGC~4636 and Section \ref{sec:x-ray} describes our method for extracting high resolution mass profiles from \Chandra\ X-ray data. In Section \ref{sec:results} we present our results and comparison to the GC analysis of \citet{chakrabarty08}, and in Section \ref{sec:discuss} we discuss the implications of our results.
\section{NGC~4636}
\label{sec:N4636}
We have chosen to explore the properties of the galaxy NGC~4636 through a detailed X-ray mass analysis and comparison to GC data. It is a particularly interesting target, as the observed bubbles and shocks seen in the \Chandra\ data \citep{jones02,osullivan05b} suggest departures from hydrostatic equilibrium in the galaxy core. The availability of dynamical data allows the plausibility of the assumption of hydrostatic equilibrium to be explored, and the systematics of the analysis methods to be investigated.

NGC~4636 is situated in a group at the edge of the Virgo cluster \citep{nolthenius93}, of which it is the brightest group galaxy \citep{osmond04}. The group is dynamically mature, and has a virial mass of 3.1$\pm$1.1$\times$10$^{13}$M$_{\odot}$ \citep{brough06b}. In studies of the X-ray properties of the group, \Rosat\ PSPC observations have shown the galaxy to have extended X-ray emission, reaching far beyond the optical limit of the galaxy \citep{trinchieri94}. The central regions of the galaxy have been studied in detail using \Chandra\ data by \citet{jones02}, who identified the presence of shocks caused by recent AGN activity, and by \citet{osullivan05b}, who found evidence for gas mixing. Analysis of \XMM\ data by \citet{finoguenov06} found features that include a plume of low entropy gas approximately 10$^{\prime}$ from the centre of the system, interpreted by the authors as evidence of stripping. The X-ray mass profile of the system has been previously studied by \citet{loewenstein03}, who found an enclosed mass at 35\,kpc of $\sim$ 1.5$\times$10$^{12}$M$_{\odot}$, strong evidence for a massive dark matter halo in this system. NGC~4636 also hosts a powerful AGN, with a radio power of log $L_{1.4GHz}$ = 21.79 \citep{jetha07}.

Following \citet{chakrabarty08} (hereafter referred to as CR08) we assume a distance of 16\,Mpc for NGC~4636 throughout. The location, scale and basic properties of NGC~4636 are shown in Table \ref{tab:props}. Right ascension, declination and redshift are from the NASA/IPAC Extragalactic Database (NED)\footnote{http://nedwww.ipac.caltech.edu/}. The $K$-band luminosity was calculated from the 2MASS $K$-band magnitude (Skrutskie et al. 2006), and the $B$-band magnitude was calculated from the extinction corrected $B_{T}$ magnitude from HyperLeda \citep{paturel03}. The effective radius ($R_{eff}$) is the radius enclosing half the light from the galaxy, and $D_{25}$ is the diameter of the isophote describing a surface brightness of 25~mag~arcsec$^{-2}$; both of these parameters are quoted here for the $B$-band \citep{RC3}. 

\begin{table}
\begin{center}
\caption{Location, scale and basic properties of NGC~4636\label{tab:props}}
\begin{tabular}{lcr}
\tableline\tableline
  R.A. (J2000.0) & & 12$^{h}$42$^{d}$49.9$^{s}$\\
  Dec. (J2000.0) & & +02$\arcdeg$41$\arcmin$16$\arcsec$\\
  Redshift & & 0.003129\\
  Distance & & 16~Mpc\\
  1 arcmin = & & 4.7~kpc\\
  log $L_{B}$ ($L_{B,\odot}$) & & 10.47\\
  log $L_{K}$ ($L_{K,\odot}$) & & 11.11\\
  $R_{eff}$ ($B$-band)$^{\dagger}$ & & 1.48$^{\prime}$\\
  $D_{25}$ ($B$-band)$^{\dagger}$ & & 6.03$^{\prime}$\\
  log $L_{x}$ (ergs$^{-1}$)$^{\ddagger}$ & & 41.59\\
  \tableline
  \end{tabular}
\tablenotetext{$^{\dagger}$}{RC3 \citep{RC3}}
\tablenotetext{$^{\ddagger}$}{\citet{osullivan01b}}
\end{center}
\end{table}
\section{X-ray data analysis}
\label{sec:x-ray}
Our aim in this analysis is to produce a high resolution mass profile from \Chandra\ data. We have achieved this through a two stage approach, by firstly concentrating on constraining the temperature profile, followed by determining the gas density profile in much greater detail. This is a similar technique to that employed by \citet{vikhlinin06}, although here we use the \XSPEC\ \textsc{projct} model to deproject the spectra. The full analysis procedure is described in detail below, and this procedure has already been applied to the early-type galaxy NGC 1407 \citep{romanowsky08}.

\subsection{Initial data reduction}
\label{sec:init}
NGC~4636 was observed by \Chandra\ on February 14th 2003 (obs ID = 3926) for 75.69~ks, and we used this archival data in the following. The initial data reduction was performed using version 3.4 of the \Chandra\ Interactive Analysis of Observations\footnote{http://cxc.harvard.edu/ciao} (\ciao) with \caldb\ version 3.4.2. We extracted a new level 2 events file from the level 1 events file, and removed events with \ASCA\ grades of 1, 5 and 7. Bad pixels were also removed from the analysis, and the appropriate gain file and time-dependent gain correction were applied. Flares were eliminated from the events file by extracting a light curve from each of the back-illuminated chips, and one from the front-illuminated chips. The light curves were filtered using the `lc\_clean' script of Markevitch\footnote{http://cxc.harvard.edu/ciao3.4/ahelp/lc\_clean.html}, resulting in a cleaned exposure time of 74.7~ks.

Point sources were detected using the \textsc{ciao} tool `wavdetect', and were excluded from further analysis. Spectra and response files were extracted from the cleaned events file, following the \textsc{ciao} analysis threads. The blank sky background files\footnote{http://cxc.harvard.edu/contrib/maxim/acisbg/} were used to extract background spectra; this was done over the same area as the source regions. The background spectra were normalised at high energies (9--12~keV) to match the source spectra. The use of blank sky backgrounds has been questioned in analyses of diffuse emission \citep[e.g.][]{humphrey06}, but this is particularly a problem in systems or regions of low surface brightness, where decomposing the diffuse emission and the background components can be very difficult. In the case of NGC~4636, the observation is heavily source dominated, so this is not a significant issue. However, we investigate the implications of using the blank sky backgrounds on our analysis in Section \ref{sec:blank}.

\begin{figure}
\includegraphics[width=8cm]{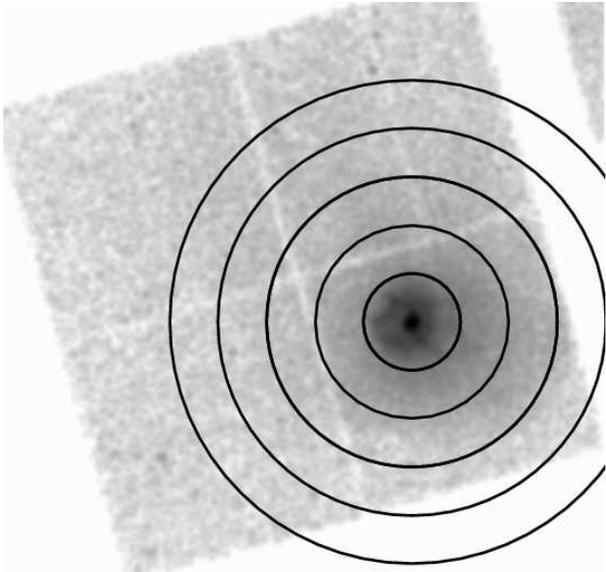}
\caption{A smoothed \Chandra\ image of NGC 4636, extracted across the energy range 0.3--2.0~keV. The circles show radii of 100$^{\prime\prime}$, 200$^{\prime\prime}$, 300$^{\prime\prime}$, 400$^{\prime\prime}$ and 500$^{\prime\prime}$ for reference, centred on the galaxy co-ordinates from NED.}
\label{fig:smooth}
\end{figure}

To indicate the scale of the disturbances in the centre of NGC 4636, Figure \ref{fig:smooth} shows an image extracted from the \Chandra\ data, across the energy range 0.3--2.0~keV, and smoothed using the \textsc{ciao} tool `aconvolve'.

\subsection{Spectral analysis}
Our two-stage spectral analysis is designed to extract high resolution mass profiles. We initially extract spectra from a series of wide concentric annuli, which we term the \textit{coarse} stage, followed by extracting spectra from much thinner annuli during the \textit{fine} stage. The coarse stage robustly constrains the temperature, and the fine stage incorporates these constraints in determining the gas density. The procedure is explained in detail below. We performed all the spectral analysis using \XSPEC\ Version 11.3.2t, and all spectra were fitted in the energy range 0.7--7.0~keV. 

\subsubsection{Coarse spectral analysis}
\label{sec:coarse}
Initially spectra and their associated responses were extracted from annuli centred on the galaxy co-ordinates from NED (shown in Table \ref{tab:props}), and background spectra were extracted from the blank sky backgrounds. The annuli were chosen to contain a net number of counts that allowed for both a successful deprojection of the spectra, as well as placing robust constraints on the temperature. It was found that a criterion of 8000 net counts per spectrum was more than adequate, yielding 13 radial bins. This criterion could be further relaxed and still provide a good fit to the spectra, however due to the nature of our approach, we probe the gas on a finer radial scale in the second stage of the analysis. The central 0.3$^{\prime}$ ($\sim$1.4~kpc) was excluded from the analysis, due to the sudden peak in surface brightness in the image in this region. 

We fitted absorbed APEC models in each annulus, with an additional power-law component subject to the same absorption to constrain the contribution from unresolved Low Mass X-ray Binaries (LMXBs). The addition of this model component is explained in detail in Section \ref{sec:lmxb}, but it is important to note here that this component is modelled as a background component, and is not deprojected. We fixed the absorption (\NH) at the Galactic value of 1.82$\times$10$^{20}$cm$^{-2}$ \citep{dickey90} throughout, and all abundances are quoted as those of \citet{grevesse98}. The spectra were then deprojected using the \textsc{projct} model in \XSPEC, under the assumption of spherical symmetry. During this procedure, the abundance was tied between all annuli, as otherwise it was unconstrained in some spectra. We discuss this assumption in full in Section \ref{sec:Z}. We define a characteristic radius, $r$, for each annulus using the emission-weighted calculation of \citet{mclaughlin99},
\begin{equation}
\label{eqn:r}
r = [0.5(r^{3/2}_{in} + r^{3/2}_{out})]^{2/3}
\end{equation} 
\noindent{where $\ensuremath{r_{in}}$ and $\ensuremath{r_{out}}$ are the inner and outer radial bounds of the annulus respectively. The deprojection therefore yields three-dimensional temperature, abundance and \NH\ as a function of radius.

We fit smoothing spline functions to the deprojected profiles, using the \sspline\ function from the \rproj\ statistical package \citep{Rcite}, to give a functional form for each deprojected profile. In this case the abundance and \NH\ are constant as a function of radius. The benefits of the smoothing spline function are that it responds to natural variation in the profile, without imposing a prescribed analytic form. Statistical fluctuations are limited by weighting the fit using the inverse variances of the parameters from a series of Monte Carlo realisations, which we describe in Section \ref{sec:MC}. The end product at this stage is a smooth, continuous functional form for the deprojected temperature profile; the data and associated smoothing spline function are shown in Figure \ref{fig:temp}.

\begin{figure}
\includegraphics[width=8cm]{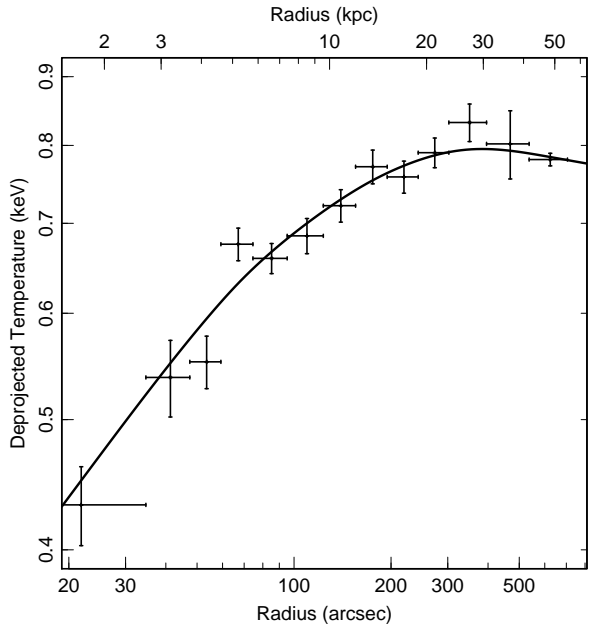}
\caption{The deprojected temperature profile for NGC~4636 and the associated smoothing spline fit, shown as a solid line (see text for details). The vertical error bars are 1$\sigma$ errors from 200 Monte Carlo realisations of the procedure and horizontal error bars show the radial extent of each annulus.}
\label{fig:temp}
\end{figure}

\subsection{Low mass X-ray binary component}
\label{sec:lmxb}
In addition to the emission from the diffuse hot gas, there is a hard X-ray contribution to the spectrum from unresolved Low Mass X-ray Binaries (LMXBs). This was modelled using a power-law model component subject to the same absorption as the APEC model, the index of which was fixed at 1.56 \citep{irwin03}. Although $\sim$50\% of the LMXBs in NGC~4636 are associated with GCs, and the light profile of GCs differs compared to the halo light profile \citep{kim06}, the \textit{distribution} of LMXBs is comparable to the halo light profile \citep{kim06}. In treating the LMXB component, we therefore make the assumption that the distribution of LMXBs follows the halo light of the galaxy, which we approximate with a de Vaucouleurs profile with $R_{eff}$ = 1.48$^{\prime}$ \citep{RC3}, equal to 6.9~kpc for our assumed distance of 16~Mpc. The de Vaucouleurs profile is an appropriate choice for galaxies with $R_{eff}$ greater than 6.3\,kpc \citep{prugniel97}. 

We fit absorbed apec+powerlaw models to the spectra in \XSPEC, to determine the normalisation of the power-law model component, under the constraint that the normalisation in each annulus should follow the overall shape of the light profile. In practice, this means that the model normalisations are tied in proportion to the shape of the light profile. We use the \XSPEC\ command \textit{fakeit} to fake a spectrum corresponding to the power-law model in each annulus, which is then added to the background spectrum. This means that this model component is not deprojected. The same approach is applied in the fine stage to quantify the LMXB component. 

The contribution from LMXBs depends on the optical luminosity of the galaxy \citep{osullivan01b,kim04}, allowing a consistency check on our adopted approach. Over the radial range covered by our coarse bins, and in the 0.5--0.7~keV energy range, the total flux from the two-dimensional spectral fitting to the coarse spectra is 9.14$\times$10$^{-12}$\,\ergps\,cm$^{-2}$. The flux from just the power-law model component is 1.49$\times$10$^{-13}$\,\ergps\,cm$^{-2}$, giving a fractional contribution to the total flux from unresolved LMXBs of 1.6\%. Assuming that the X-ray luminosity from discrete sources is log $L_{dscr}$ = 29.5\,\ergps\,$L_{B\odot}^{-1}$ \citep{osullivan01b}, we expect, given a $B$-band luminosity of 10.47 (Table \ref{tab:props}), a contribution from the discrete sources of approximately log $L_{dscr}$ = 39.97\,\ergps. Assuming log $L_{x}$ = 41.59 \citet{osullivan01b}, the expected unresolved source contribution to the total luminosity is $\sim$2\%. Our unresolved flux is reasonable given this prediction. The high resolution of \Chandra\ and the deep observation of NGC~4636 will have allowed more of the brightest point sources to be detected and excluded than in the \Rosat\ data of \citet{osullivan01b}, which could easily lead to the slight difference in the predicted and observed unresolved source fraction.

As a thermal bremsstrahlung component with a fixed temperature of 7.3\,keV can also be used to describe the LMXB spectrum \citep{irwin03}, we tested the use of this component instead of the power-law described above. There was no improvement in the fit, the fitted parameters were consistent with those recovered from using the power-law, and the percentage of the total flux in the 0.5--7.0\,keV energy range was found to be 1.5\,\%, again consistent with that recovered from the fitting using the power-law model.

\subsection{Fine spectral analysis}
\label{sec:fine}
We next determine a set of finely spaced annuli, from which source spectra, background spectra, and the appropriate response files are extracted. Our motivation here is to model only the gas density; the remaining parameters in our model are described by the fitted functions from the coarse stage. The fine annuli are spaced using a net counts criterion, however, as we are now only fitting for one parameter, the number of net counts in each annulus can be considerably reduced. We use 2000 net counts per spectrum (51 spectra) as a compromise between resolution and the time taken to perform the Monte Carlo error analysis. Under this criterion, the annular width of the bins ranges between a minimum of 2.95$^{\prime\prime}$ and a maximum of 66.9$^{\prime\prime}$, meaning that the bin width exceeds the PSF at all radii. We determined the characteristic fine radii for these annuli using Equation~\ref{eqn:r}. 

Using the functional fits to the deprojected profiles described in Section \ref{sec:coarse}, we interpolated the values of deprojected temperature, abundance and \NH\ at the characteristic fine radii. These parameters were kept fixed in the subsequent \textsc{projct} model fit to the fine spectra. The contribution from LMXBs was included as described in Section \ref{sec:lmxb}, and to speed up the fitting in this stage, we only extracted spectra across 8 channels. There is therefore just one free parameter at this stage in the deprojection --- the APEC model normalisation, \ensuremath{K}, from which the gas density can be directly determined as
\begin{equation}
K = \frac{10^{-14}}{4\pi(D_{A}(1+z))^{2}}\int~n_{H}n_{e}dV,
\end{equation}
where \ensuremath{D_{A}} is the angular diameter distance to the galaxy, \ensuremath{z} is the redshift of the galaxy, and \ensuremath{n_{H}} and \ensuremath{n_{e}} are the number density of hydrogen and electrons respectively, and we assume that \ensuremath{n_{H}}/\ensuremath{n_{e}} = 1.17. Therefore, for a particular spherical shell of volume \ensuremath{dV}, the gas density can be recovered. The resulting gas density profile is shown in Figure \ref{fig:density}, and has been fitted with a $\beta$-model, yielding parameters of r$_{core}$ = 31.6$^{\prime\prime}$ and $\beta$ = 0.5 for a reduced \chisq\ of 4.7 (48 degrees of freedom). The benefit of this two stage approach is that we retain a robust temperature profile, but improve the resolution of the gas density profile. 

\begin{figure}
\includegraphics[width=8cm]{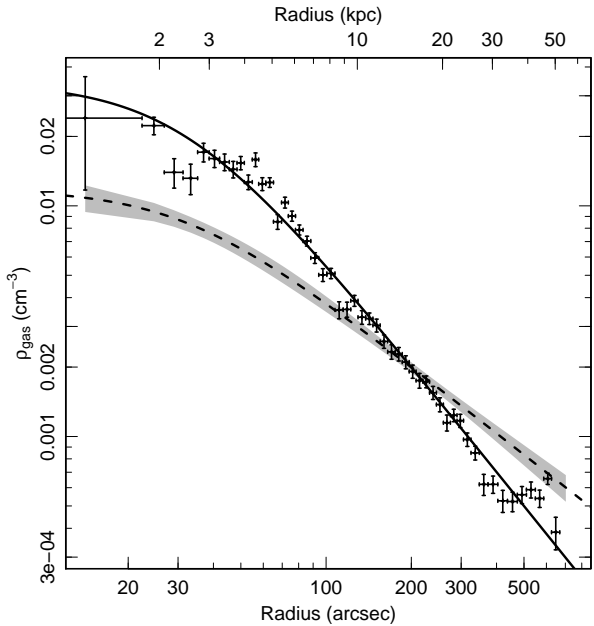}
\caption{The gas density profile for NGC~4636, determined from the finely binned spectral analysis. The solid line is a $\beta$-model fit to the data points, and the vertical error bars show the 1$\sigma$ errors from 200 Monte Carlo realisations of the procedure. Horizontal error bars show the radial extent of each annulus. The dashed line shows the effect on the gas density profile of allowing for the observed projected abundance gradient (see Section \ref{sec:Z}), and the associated 2$\sigma$ confidence region is derived from $\beta$-model fits to 100 Monte Carlo realisations.}
\label{fig:density}
\end{figure}

It is apparent from Figure \ref{fig:density} that the $\beta$-model shape is not successful in describing the shape of the gas density profile at all radii, and fluctuations from this smooth profile can be seen. However, the calculation of the resulting mass profile requires a smooth and continuous function, and we shall proceed with the use of the fitted $\beta$-model in this context. The observed fluctuations can be readily understood in terms of the observed disturbances in the X-ray emitting gas. Examining the X-ray image shows that the edge of the central shock region occurs at a radius of approximately 100$^{\prime\prime}$, which corresponds to a slight depression in the gas density profile shown in Figure \ref{fig:density}. The feature at $\sim$\,60$^{\prime\prime}$ arises as a consequence of the shocks in the galaxy core. The nature of the spectral analysis and deprojection means that the profiles shown here represent azimuthally averaged measurements, so very localised features in the hot gas would be smoothed out. In the context of the gas density behaviour beyond $\sim$500$^{\prime\prime}$, we note that \citet{trinchieri94} reported a flattening of the gas density distribution at radii of 6$^{\prime}$--8$^{\prime}$ on the basis of \Rosat\ data. We will examine this feature in further detail in Section \ref{sec:flat}.

\subsection{Monte Carlo error analysis}
\label{sec:MC}
We have employed a Monte Carlo (MC) approach to calculate the errors associated with the procedure, implemented in both the coarse and fine stages in an analogous way. Initially, the best-fitting \textsc{projct} model from the coarse analysis is used to produce a series of spectra using the \XSPEC\ command \textit{fakeit} with the inclusion of random Poisson noise. These spectra are then fitted with a \textsc{projct} model, and the process is repeated 200 times, from which the standard deviation is used to define 1$\sigma$ errors on the coarse profiles. The errors are also calculated for the fine stage of the analysis by using the best-fitting \textsc{projct} model from the fine analysis to fake a series of spectra. These are fitted with \textsc{projct} models to determine the APEC model normalisation, from which 1$\sigma$ errors are determined at the fine radii. The MC realisations of the temperature profile are fitted with the \sspline\ algorithm to determine continuous functions, which are used in conjunction with $\beta$-model fits to the MC gas density profiles to yield 200 MC realisations of the mass profile. From this suite of mass profiles, errors are estimated by determining the 1$\sigma$ spread in the functions evaluated at the characteristic radii.
\section{Results}
\label{sec:results}
Here we present our X-ray derived mass profile, before comparing this to the results derived from a dynamical analysis of the GC population, performed by CR08.

\subsection{X-ray mass profile}
The resulting fits to the temperature (Section \ref{sec:coarse}) and gas density profiles (Section \ref{sec:fine}) are used to determine the mass within a given radius \ensuremath{M(<r)}, in the following way \citep{fabricant80},
\begin{equation}
\label{eqn:mass}
M(<r) = \frac{k_{B}~Tr}{G{\mu}m_{p}}\left(-\frac{d~ln\rho}{d~lnr}-\frac{d~lnT}{d~lnr}\right),
\end{equation}
where $\rho$ is the gas density, $T$ is the temperature, $G$ is the gravitational constant, $\mu$ is the mean molecular mass (assumed here to be 0.593 for a fully ionised plasma) and $m_{p}$ is the mass of a proton. The mass profile derived from the X-ray analysis is shown in Figure \ref{fig:mass}. The confidence region shows the 2$\sigma$ spread from the 200 Monte Carlo realisations of the mass profile. The X-ray gas density and temperature measurements within 30$^{\prime\prime}$ are well-constrained in the spectral fits, and are not obviously biased. However, the calculation of the mass profile yields an unphysical negative mass in this region, indicating the requirement for additional non-thermal pressure support within the central $\sim$\,3\kpc. 

Performing a mass analysis on an earlier \Chandra\ dataset, \citet{loewenstein03} determined a total mass of $\sim$1.5$\times$10$^{12}$M$_{\odot}$ at $\sim$35\,\kpc. Correcting for the different assumed distance, we plot the enclosed mass recovered by \citet{loewenstein03} in Figure \ref{fig:mass}. This falls just below the 2\,$\sigma$ confidence bound on our original mass profile. \citet{loewenstein03} also found the mass to increase as $r^{1.2}$ over the radial range studied (0.7--35\,\kpc). Fitting a powerlaw model to our data, we find the same slope (1.19$\pm$0.01) outside 5\,\kpc. Within this radius, our mass profile falls away more steeply, a consequence of the steeper temperature profile recovered from our analysis in the inner regions.

Figure \ref{fig:rhotot} shows the total mass density profile from the X-ray procedure evaluated at the finely spaced radii, and the total mass density profile recovered when the abundance gradient is included in the fitting as detailed in Section \ref{sec:Z}. The mass density profile of elliptical galaxies is the combination of the stellar mass density and the underlying dark matter density, and the stellar mass density dominates within approximately 1$R_{eff}$ \citep{mamon05a}.

\begin{figure}
\includegraphics[width=8cm]{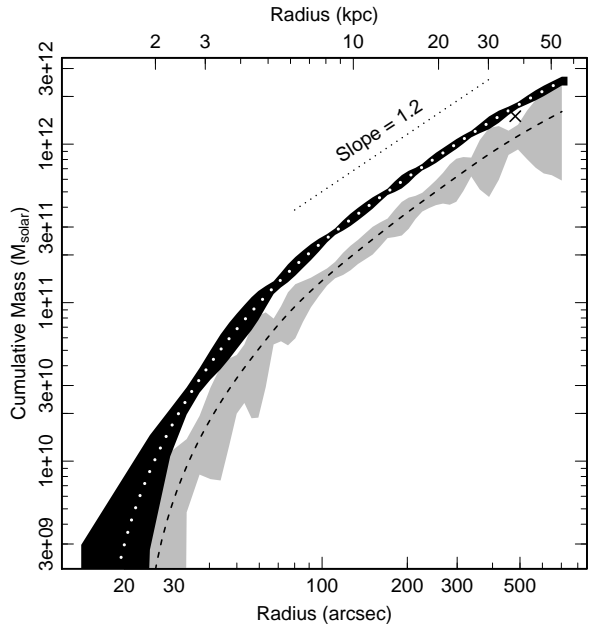}
\caption{The cumulative mass profile of NGC~4636. The white dotted line and the associated black (2\,$\sigma$) confidence region show the results of our X-ray mass analysis. The confidence region has been determined from 200 Monte Carlo realisations of our procedure. The dashed line shows the effect on the calculated mass of allowing for the abundance gradient (see Section \ref{sec:Z}), and the associated 1$\sigma$ confidence region shows the results of 100 Monte Carlo realisations of the procedure. The black dotted line shows the powerlaw slope of $r^{1.2}$ determined by \citet{loewenstein03}, with arbitrary normalisation, and the cross point shows the total mass measured by \citet{loewenstein03} at 35\,kpc (corrected for our assumed distance). }
\label{fig:mass}
\end{figure}
\begin{figure}[t]
\includegraphics[width=8cm]{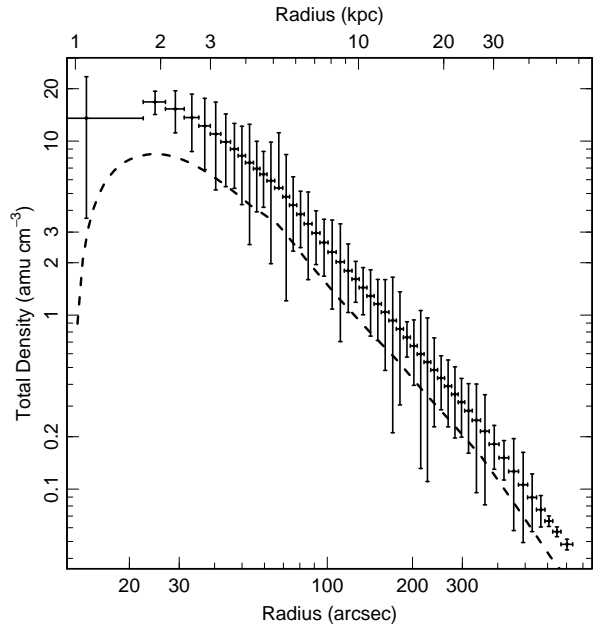}
\caption{The total mass density profile of NGC~4636 derived from the cumulative mass profile shown in Figure \ref{fig:mass}, and evaluated at the characteristic fine radii. The horizontal error bars show the radial width of each bin, and the vertical error bars show the 1$\sigma$ spread of 200 Monte Carlo realisations evaluated at each radius. The dashed line shows the total mass density profile when an allowance is made for the observed abundance gradient (see Section \ref{sec:Z} for details).}
\label{fig:rhotot}
\end{figure}
\subsubsection{Implications of allowing for the abundance gradient}
\label{sec:Z}
Recent work by \citet{rasmussen07} has shown abundance gradients to be prevalent in galaxy groups, and we now consider the implications of implicitly assuming a flat abundance profile, when the projected analysis reveals an abundance gradient (Figure \ref{fig:Z}). In the innermost bins, the abundance is poorly constrained and reaches the default \XSPEC\ fitting limits --- this was our main motivation for imposing a flat profile. This is probably a consequence of multiple temperature and abundance components in the inner regions (see Section \ref{sec:multi}). The fitted deprojected abundance is also shown in Figure \ref{fig:Z}, and it can clearly be seen that imposing this criterion underestimates the abundance within 200$^{\prime\prime}$ and overestimates the abundance outside this radius. We are motivated to test the effect on the mass profile of assuming a constant abundance as this is often employed to satisfactorily constrain model parameters in less luminous systems, or in cases of poorer data quality, and the effect of such an assumption has not been studied. It is a particularly important issue in low temperature systems where line emission dominates. 

We repeated our coarse deprojection, fixing the deprojected abundances at their projected values. It is very likely therefore, that we have now \textit{overestimated} the very central abundance due to its poor constraints, so we consider the following to be an upper limit on the effects of allowing for the abundance gradient. The abundance and APEC model normalisation play off against each other due to line emission dominating the flux at low temperatures, so to see the full effects on the gas density we proceeded with the fine stage. We fixed the temperature profile in the fine stage deprojection at the values interpolated from the original fit to the data (see Section \ref{sec:coarse}), as the variation in temperature caused by allowing for the abundance gradient was well within the 1\,$\sigma$ errors of the original temperature profile. To establish the errors in this analysis, we performed 100 Monte Carlo realisations of the procedure, using the MC realisations from the coarse stage to weight the smoothing spline fit to the abundance profile at the beginning of the fine stage. 

The effect of allowing for the abundance gradient is to flatten the gas density at all radii. Fitting a $\beta$-model, weighted by the inverse variance from the MC realisations, gives $\beta$ = 0.3 and $r_{core}$ = 27.2$^{\prime\prime}$ and is shown in Figure \ref{fig:density}. The associated 1\,$\sigma$ confidence region shows the range of $\beta$-model fits allowed by the MC realisations. The subsequent effects on the mass profile and total density profile are shown in Figures \ref{fig:mass} and \ref{fig:rhotot}. The effect of the abundance gradient is to reduce the mass at all radii by a factor of $\sim$\,1.6. This demonstrates the intricacies involved in the detailed X-ray analysis, as this effect is not allowed for by our Monte Carlo procedure, which also ties the abundances in the deprojected fit and hence this is a key systematic in the application of our method.  
\begin{figure}
\includegraphics[width=8cm]{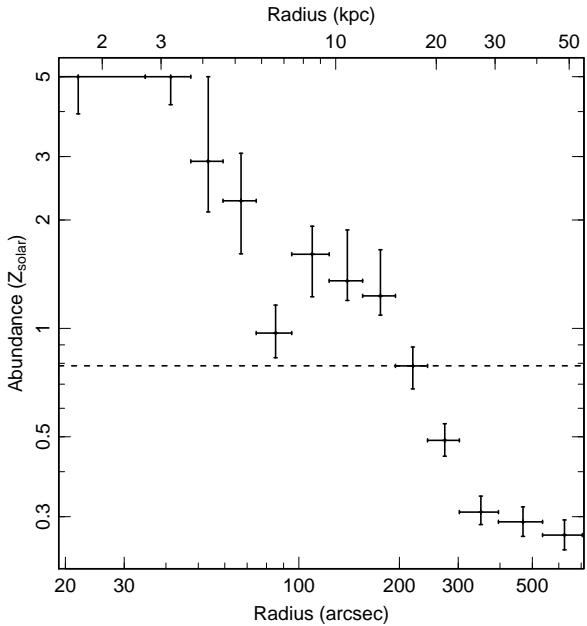}
\caption{The projected abundance profile of NGC~4636 (data points), with errors estimated from \XSPEC. In the two innermost radial bins, the abundance is unconstrained, shown by the data points reaching the \XSPEC\ default fitting limit of 5.0$Z_{\odot}$. The dashed line shows the fitted abundance in the deprojection, where the abundance is tied between all the radial bins.}
\label{fig:Z}
\end{figure}
\subsubsection{Implications of assumed single temperature model}
\label{sec:multi}
Throughout our analysis we have implicitly assumed that the gas in each annulus is single phase. If the gas is multi-phase in these regions, the recovered abundances may be affected by ``Fe-bias'' \citep{buote00a}, where the abundance of multi-phase gas is underestimated if a single temperature model is fitted. The disturbed nature of the gas in NGC~4636 \citep{jones02,osullivan05b} suggests multiple temperatures and abundance will be present in each coarse region. Such an integrated spectrum would have a broader iron peak, and would require very high quality spectra to separate the individual components. Using the 2-d coarse spectra, we tested the addition of an extra APEC model component, but found that this did not improve the fit at any radius. Although this system has been shown to host cavities \citep{ohto03} and also shows surface brightness features \citep{osullivan05b}, it appears that when considering an azimuthally averaged profile with a sufficiently large number of counts, a single temperature model is acceptable. In terms of our mass analysis, it is important to have a good representation of the temperature profile, even if the model itself does not give the most statistically accurate fit.

\subsubsection{Implications of using blank sky backgrounds}
\label{sec:blank}
To test the sensitivity of our results to the use of the blank sky backgrounds, we performed the following tests. Using the outermost coarse annulus, which will be the most sensitive to the background, we fitted a simple absorbed APEC model with \NH\ fixed at the Galactic value, to recover the temperature, abundance and APEC model normalisation shown in Table \ref{tab:back}. We scaled the exposure time of the blank sky background spectrum for this annulus up and down by 10 percent, re-fitting each time to effectively alter the background normalisation, as the same number of counts are collected over a differing time period. The recovered parameters are also shown in Table \ref{tab:back}. We find that the fitted parameters for the increased exposure time are consistent within 1\,$\sigma$, as is the recovered temperature for the decreased exposure time case, with the abundance and model normalisation consistent with the original fit within 2\,$\sigma$. These tests indicate that our results are not sensitive to variations in the background level at the level of 10 percent, and emphasises how robust the temperature measurement is at these low temperatures due to the dominance of the line emission.

\begin{table}
\begin{center}
\caption{The recovered parameters from tests carried out on the use of the blank sky backgrounds.\label{tab:back}}
\begin{tabular}{cccc}
\tableline\tableline
& kT & Abundance & APEC norm\\
& (keV) & (Z$_{\odot}$) & \\
\tableline
Blank sky & 0.79$\pm$0.01 & 0.17$^{+0.03}_{-0.02}$ & 1.26$^{+0.12}_{-0.11}$~$\times$10$^{-3}$\\
Blank sky + 10\% & 0.79$\pm$0.01 & 0.15$\pm$0.02 & 1.43$^{+0.12}_{-0.11}$~$\times$10$^{-3}$\\
Blank sky - 10\% & 0.79$\pm$0.01 & 0.21$^{+0.04}_{-0.03}$ & 1.06$^{+0.11}_{-0.12}$~$\times$10$^{-3}$\\
\tableline
\end{tabular}
  \tablenotetext{}{\textsc{Notes:} The fitted model parameters are shown for an absorbed APEC model fit to the outermost coarse annulus, for the assumed blank sky background, and also for variations of $\pm$10\% in the normalisation of this background. Errors are derived from \XSPEC\ and are 1$\sigma$.}
\end{center}
\end{table}

\subsubsection{Implications of assumed LMXB model}
The $K$-band is a better description of the older stellar populations of early-type galaxies than the $B$-band, and therefore may better describe the LMXB distribution. We considered the effects of assuming a de Vaucouleurs profile for the LMXB population with a $K$-band $R_{eff}$ of 56.2$^{\prime\prime}$ \citep{jarrett03}. This assumption reduces the gas density at all radii by approximately 3\,\%, and makes no discernible change to the innermost nine temperature points (less than 1\%). Instability from the deprojection procedure is however visible at the largest radii. The fitted abundance, which was again tied between the annuli, was 0.85$Z_{\odot}$ using the $K$-band $R_{eff}$, compared to 0.79$Z_{\odot}$ in the original analysis. We therefore assert that the effect on the mass profile of the treatment of any abundance gradient is more crucial in this case than the intricacies of the treatment of the LMXBs. This may not be the case for galaxies where the unresolved source emission is a higher fraction of the overall X-ray emission.

\subsection{Comparison to dynamical mass estimate}
\label{sec:dynmass}
A sample of 174 GCs in NGC~4636 were tracked for their line-of-sight velocities in the observational programme of \citet{dirsch05} and were used to assess the mass profile of NGC~4636 by \citet{schuberth06}. CR08 input these kinematic data into the Bayesian non-parametric algorithm CHASSIS \citep{chakrabarty01}. This invokes a Markov Chain Monte Carlo (MCMC) optimiser to recover the most likely equilibrium distribution function from which this kinematic data could have been drawn, given the recovered potential in which the sample of GCs resides. This potential is expected to be the gravitational potential of the galaxy itself, from which the total (luminous+dark) matter density of NGC~4636 is estimated. Motivated by a desire to understand and test the underlying assumptions of these two independent methods, we can view the X-ray mass profile in comparison to the dynamical estimate of the total mass distribution. At this point, it merits mention that in its current form, CHASSIS assumes isotropy in phase space, although work is underway to relax the requirement of velocity isotropy (Chakrabarty $\&$ Saha, in preparation).

\begin{figure}
\includegraphics[width=8cm]{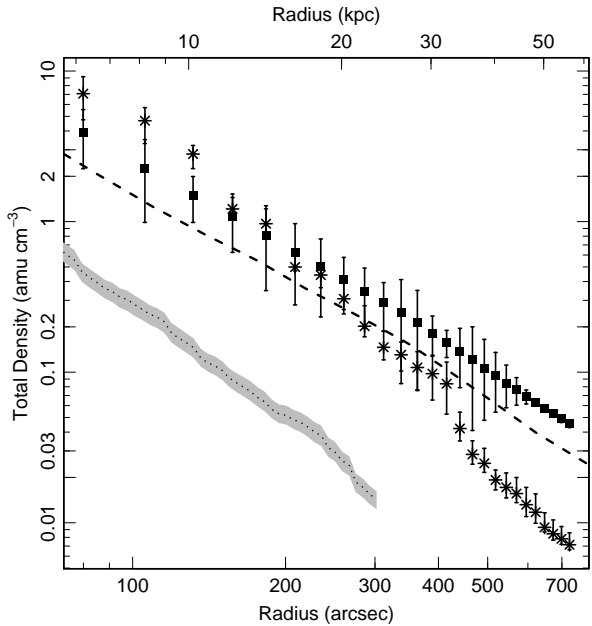}
\caption{The total mass density profile of NGC~4636. Solid squares show the results from the X-ray analysis presented here, with vertical error bars showing 1$\sigma$ errors from 200 Monte Carlo realisations of the procedure. Stars show the results from the GC analysis of CR08, where vertical error bars are the 1$\sigma$ spread in mass models derived from the CHASSIS algorithm. Note that the radial range of the X-ray measurements has been restricted to that determined by the dynamical mass measurements, and these radii have been used to evaluate the X-ray profile. The dashed line shows the recovered mass density profile when the abundances are fixed at their projected values (see Section \ref{sec:Z}). The dotted line and confidence region show the estimated stellar mass density (see text for details). }
\label{fig:compare}
\end{figure}

Figure \ref{fig:compare} shows the total mass density profile of NGC~4636 (star symbols) recovered in RUN~I of CHASSIS by CR08; we refer the reader to this work for more information. The mass density from the X-ray analysis is shown as solid squares, and has been evaluated at the radii of the dynamical mass profile. The errors on the dynamical mass estimate indicate the $\pm$1$\sigma$ spread in the mass models about the most likely configuration, as determined by the MCMC optimiser that is used in CHASSIS. We show the stellar mass density determined from the $K$-band luminosity density profile presented by CR08, assuming a stellar mass-to-light ratio in the $K$-band of 0.83. This is a colour-dependent estimate using the total $B - V$ colour from HyperLeda\footnote{http://leda.univ-lyon1.fr/} of 0.94, and converting to the $K$-band mass-to-light ratio following the prescription of \citet{bell03}. 

The general nature of the dynamical and X-ray mass distributions is similar to about 30 kpc, beyond which, the X-ray mass exceeds the dynamical mass, by a factor of $\sim$4.5 at 40 kpc. There is also an indication of a `break' in the dynamical mass density profile, at a little over 30~kpc. Of the 174 GCs studied by \citet{schuberth06}, only 15 of these are at radii greater than 7.5$^{\prime}$. In terms of the GC density distribution, a steepening is observed between approximately 6$^{\prime}$ and 8$^{\prime}$ \citep{dirsch05}, noted by \citet{schuberth06} to be inconsistent with NGC~4636 being in a dark matter potential which smoothly reaches to large radius. We will return to this point in Section \ref{sec:discuss}. The key difference between the profiles at large radii is the shape; the GC profile appears to `break' at approximately 400$^{\prime\prime}$, dropping away more steeply than the X-ray derived profiles. We note that the effect of allowing for the observed abundance gradient in the X-ray analysis reduces the mass density at all radii, improving the agreement in the outer regions.

CR08 fit a \citet{navarro96} density profile, hereafter NFW profile, to the total density outside $\sim$32\kpc\ recovered from the GC analysis using CHASSIS. They find a concentration of 9, and scale radius $r_{s}$ of 33.7\kpc$\pm$11 percent. We fit the X-ray mass density profile across the radial range shown in Figure \ref{fig:compare} using the \rproj\ non-linear least squares algorithm `nls', weighting each point by its inverse variance. We find a concentration of 20.1$\pm$0.8 and a scale radius of 21.8$\pm$0.9\kpc, where the quoted errors are 1\,$\sigma$ standard errors on the fit. However, this concentration is an over-estimate due to ignoring the stellar contribution to the mass density, so we proceed to fit an NFW profile to the mass density, having subtracted the stellar mass density shown in Figure \ref{fig:compare}. There is some uncertainty in this approach due to our assumed stellar mass-to-light ratio, but this fit does recover a lower concentration (18.0$\pm$0.6), with a scale radius of 24.6$\pm$0.9\kpc, leading to an estimate for $r_{200}$ of approximately 443\kpc. This concentration is similar to NGC~720 and NGC~1407 \citep{buote07}, which are slightly cooler and warmer \citep[$\sim$0.5\keV\ and $\sim$1.0\keV;][]{osmond04} than NGC~4636 respectively. Further increasing the stellar component by increasing the mass-to-light ratio would further reduce the recovered concentration. 

If we instead fit the NFW profile to the X-ray profile where the abundance gradient has been incorporated, we find a concentration of 14.4$\pm$0.4 when we first subtract the stellar mass, with a scale radius of 26.1$\pm$1\kpc\ respectively. We show the NFW fits to the X-ray analysis when the stellar mass is subtracted in Figure \ref{fig:nfw}. These solutions also both fit on the c--M relation of \citet{buote07}. The shape of the NFW profile is a good approximation to the shape of the recovered total mass density profile, and it would be difficult to fit an NFW profile across the whole radial range to the GC data, due to the small `break' in the profile at radii of $\sim$400--500$^{\prime\prime}$. 
\begin{figure}
\includegraphics[width=8cm]{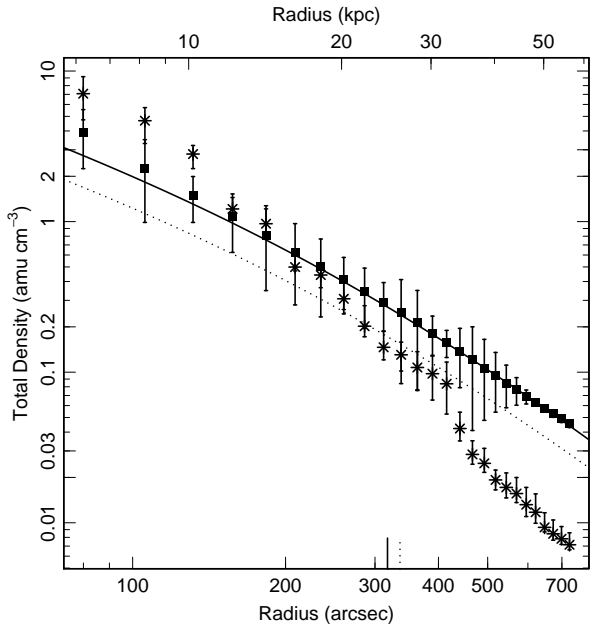}
\caption{The total mass density profile for NGC~4636. The data points are as shown in Figure \ref{fig:compare}. The solid line is the NFW fit to the main X-ray analysis, having subtracted the stellar mass component shown in Figure \ref{fig:compare}. The dotted line is the NFW fit to the X-ray analysis when the abundance gradient has been allowed for, and again the stellar mass component was subtracted to perform the fit. The short, vertical solid and dotted lines on the x axis are the scale radii of the solid and dotted fits, respectively.}
\label{fig:nfw}
\end{figure}
\subsubsection{Comparison of gravitational potential}
\begin{figure}[t]
\includegraphics[width=8cm]{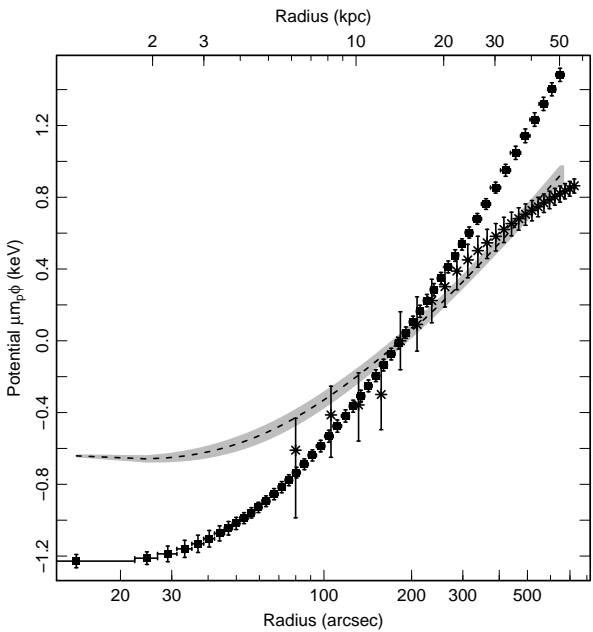}
\caption{The gravitational potential, in units of keV \citep[see, e.g.][]{churazov08}, as a function of radius recovered from the X-ray analysis using Equation \ref{eqn:pot} (solid squares), with 1\,$\sigma$ error bars from 100 Monte Carlo realisations of the method. For comparison we show the gravitational potential recovered from the GC analysis of CR08 (stars), complete with 1\,$\sigma$ error bars. The reference radius has been set (see text) such that the potential is zero at 14.2\,kpc. The dashed line and associated 1\,$\sigma$ confidence region shows the result of allowing for the abundance gradient in the analysis (see Section \ref{sec:Z} for details). }
\label{fig:pot}
\end{figure}
Perhaps a more appropriate method of comparison is to look at the gravitational potential recovered from each method. As shown by \citet{churazov08}, the gravitational potential can be easily recovered from an X-ray analysis of this type, if hydrostatic equilibrium and spherical symmetry are further assumed. The requirement that the thermal gas pressure is the only contributor to the overall pressure, yields the following expression \citep{churazov08} for the potential $\phi_{X-ray}$,
\begin{equation}
\label{eqn:pot}
\phi_{X-ray} = -\frac{k_B}{\mu~m_p}\left[\int~T\frac{~d~ln~\rho}{dr}~dr + T\right] + C,
\end{equation} 
where $C$ is an arbitrary constant and all other terms are defined as in Equation \ref{eqn:mass}. This can be directly compared to the recovered potential from the dynamical mass analysis, which has been determined in a Bayesian manner. To make the comparison, we need to assign a reference radius, at which the potential from both methods is set to zero. The choice of this radius is in fact arbitrary, but we have taken into account the properties of each profile in setting this radius to allow a useful comparison of the profiles. The shock regions in NGC~4636 extend to 100$^{\prime\prime}$ ($\sim$7.8\,kpc), and within this radius it is unlikely that the assumption of hydrostatic equilibrium is an adequate description of the state of the gas. Therefore, we set the reference radius to be 14.2\,kpc ($\sim$\,183$^{\prime\prime}$), which also takes into account the presence of close to 60 GCs within this radius, meaning that the potential from the dynamical mass analysis should be well-constrained.

We compare the potential profiles recovered from each method in Figure \ref{fig:pot}. As expected from the comparisons of the total mass density shown in Figure \ref{fig:compare}, the potential recovered from the X-ray analysis exceeds that from the GC analysis outside $\sim$\,30\,kpc. Again, the \textit{shape} of the two profiles also differs outside a radius of 30\,kpc; the X-ray gas and GCs trace the same underlying potential, so this is problematical. For comparison, we also show the results of allowing for the abundance gradient (shown as a dashed line), as explained in Section \ref{sec:Z}. Figure \ref{fig:pot} shows that in the radial range $\sim$100--300$^{\prime\prime}$, the X-ray and dynamically derived potential profiles agree within the 1\,$\sigma$ errors. Allowing for the metallicity gradient in the X-ray data appears to improve the agreement at large radius, making the X-ray and dynamical profiles consistent within the quoted 1\,$\sigma$ errors. However, care must be taken in the interpretation of these results, as the profiles have been normalised to equal zero at the same radius, and the agreement weakens if normalised elsewhere. The key point is that the shape of the X-ray and dynamically derived profiles agrees within the 1$\sigma$ errors over the radial range $\sim$\,100$^{\prime\prime}$ to $\sim$\,300$^{\prime\prime}$, but outside this radius, the gradient of the dynamical profile lessens with radius compared to the gradient of the X-ray potential profile.

\citet{churazov08} explain in detail how directly comparing the potential recovered from each method lends some insight into the magnitude of any non-thermal pressure effects, as any non-thermal pressure support in the X-ray gas would lead to a smaller change in the X-ray potential compared to the dynamical potential. Hence, the gradient of a linear fit to Figure \ref{fig:pot2} would yield information about the fractional contribution from non-thermal pressure support. However, we can immediately see that in this case there is not a simple linear expression linking the potentials, and the gradient increases with increasing potential. Following the prescription of \citet{churazov08}, we cannot attribute the behaviour at large radius (potential) to the presence of non-thermal pressure support, as this would reduce the X-ray derived potential in relation to the dynamical potential. The radial coverage of the GC data also limits the usefulness of this comparison for determining any non-thermal pressure support component in the shocked region ($<$ 100$^{\prime\prime}$).
\begin{figure}
\includegraphics[width=8cm]{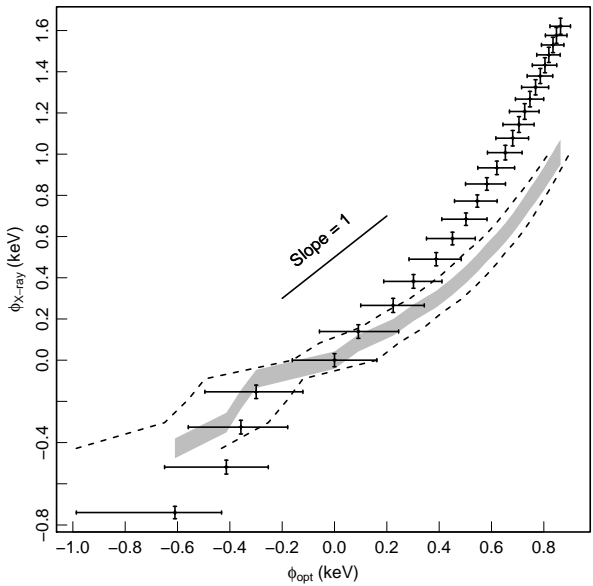}
\caption{A direct comparison of the gravitational potential recovered from the X-ray (y axis) and GC (x axis) analyses, evaluated at the radii of the dynamical profile (error bars). The grey confidence region and dashed lines show the effects of allowing for the abundance gradient in the analysis. The confidence region shows the 1\,$\sigma$ errors in the x direction, whereas the dashed lines show the 1\,$\sigma$ errors in the y direction. The solid line is for reference and shows a slope of 1.}
\label{fig:pot2}
\end{figure}

\subsubsection{Mass-to-light ratio}
We can examine the central regions in detail, by comparing the recovered $K$-band mass-to-light ratios from the two methods. The enclosed light profile was deprojected by CR08 from a $K$-band surface brightness profile provided by Tom Jarrett (see CR08 for details) from the 2MASS Large Galaxy Atlas \citep{jarrett03}. We use this enclosed light profile to determine the mass-to-light ratio from our X-ray mass profile. The comparison is shown in Figure \ref{fig:mtol}, where the mass-to-light ratio from the X-ray analysis (solid line and associated confidence region) has been capped at the limit of the light profile data (23.5\,\kpc). Within 1~$R_{eff}$, the $K$-band mass-to-light ratio derived from the X-ray analysis decreases inwards implying $M_{\ast}/L_{K}$ $<$ 3; this is the region where the stellar mass component dominates over the dark matter component \citep{mamon05a}. The mean stellar mass-to-light ratio in the $K$-band was found from 2dFGRS and 2MASS data to be 0.73~$M_{\odot,K}/L_{\odot,K}$ assuming a Kennicutt IMF, and 1.32~$M_{\odot,K}/L_{\odot,K}$ assuming a Salpeter IMF \citep{cole01}. This considered both early and late-type galaxies, but as the near-IR luminosity traces the older stellar population, it is reasonable to compare with this result. We also show the colour-dependent estimate of $K$-band stellar mass-to-light ratio from the prescription of \citet[][see Section \ref{sec:dynmass} for more details]{bell03}. \citet{humphrey06} measured stellar mass-to-light ratios for 7 early-type galaxies, both from stellar population synthesis models and from modelling X-ray derived mass profiles with dark matter and stellar components. They find stellar mass-to-light ratios ranging between $\sim$0.5 and $\sim$1.2 from the X-ray mass modelling. The stellar population models recover slightly higher values ($\sim$0.4 to $\sim$1.9), depending on the assumed IMF. This indicates that the recovered $K$-band mass-to-light ratio in the central regions is consistent with previous results, although the values at radii $<$ 30$^{\prime\prime}$ fall below the mean value of \citet{cole01} and the colour-dependent estimate of \citet{bell03}. 

One possible explanation for the low stellar mass-to-light ratios seen in the central regions from the X-ray analysis is that in this region, we are underestimating the galaxy mass. We will explore this possibility in Section \ref{sec:discuss}. However, it is clear from Figure \ref{fig:mtol} that outside 1$R_{eff}$ there is a contributive mass component in addition to that expected from the stars alone, and this is predicted by \textit{both} the X-ray and GC analyses. 
\begin{figure}
\includegraphics[width=8cm]{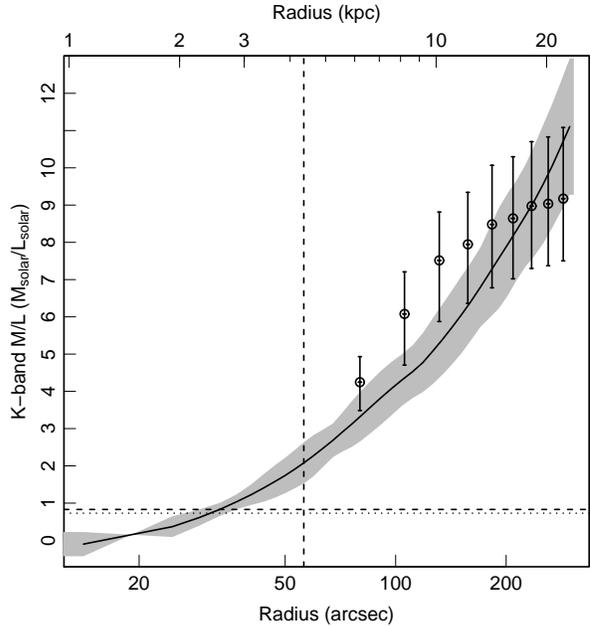}
\caption{The $K$-band mass-to-light ratio from the GC analysis of CR08 (shown as open circles with error bars) and the $K$-band mass-to-light ratio derived from the X-ray analysis (shown as the solid line). The shaded region shows the combined errors from the light profile and the X-ray mass profile. The vertical dashed line shows the $K$-band $R_{eff}$ of the galaxy from the 2MASS Large Galaxy Atlas \citep[$R_{eff,K}$ = 56.2$^{\prime\prime}$,][]{jarrett03}, the horizontal dotted line shows the mean stellar mass-to-light ratio of \citet{cole01} (Kennicutt IMF) and the horizontal dashed line shows the colour-dependent $M/L_{K}$ of \citet{bell03}. }
\label{fig:mtol}
\end{figure}
%
\section{Discussion}
\label{sec:discuss}
We have shown in Figure \ref{fig:compare} the derived total mass density profile from our X-ray analysis, and the results of the dynamical analysis by CR08 of the GC system of NGC~4636. We have also compared the gravitational potential recovered from each method, and we note the following:
\begin{enumerate}
\item Within $\sim$10~kpc, the mass derived from the dynamical analysis of CR08 exceeds the mass recovered from the X-ray analysis.
\item Between $\sim$10~kpc and $\sim$30~kpc, the profiles are consistent within the quoted 1$\sigma$ errors.
\item Perhaps most crucially, outside $\sim$30~kpc, the mass recovered from the X-ray analysis significantly exceeds that derived from the dynamical analysis.\end{enumerate}

Considering the mass-to-light profile of the system suggests a significant dark matter component outside one effective radius, independent of the analysis method. With the aim of understanding the behaviour in the inner and outer regions, we now review the observed discrepancies in terms of the key systematics of each analysis method.

\subsection{Anisotropy $\&$ CHASSIS}
The assumption that prevails within the current form of the dynamical analysis --- the algorithm CHASSIS --- is that of isotropy in phase space, and we should examine the recovered dynamical density profile (Figure~\ref{fig:compare}), in light of this assumption, or in particular, the assumption of velocity isotropy. At the outset, we note that deviation from isotropy in the true velocity space configuration of the system would urge CHASSIS to \textit{over-estimate} the mass density, at radii where anisotropy prevails \citep{chakrabarty05, chakrabarty06}. If in reality, velocity anisotropy describes the phase space distribution from which the measured GC kinematic data are drawn, then the recovered mass density distribution would be spuriously enhanced in amplitude at these radii. In other words, the ``true'' mass density would be even \textit{lower} than that indicated by the current dynamical estimates (see Figure~\ref{fig:compare}). Therefore, invoking velocity anisotropy does not help to reconcile the X-ray and dynamical mass density profiles in the outer parts of NGC~4636.

This however poses the question of whether the density distribution of CR08 is an overestimate, due to mistaking the velocity space configuration as isotropic, and if so, can we quantify how bad the assumption of isotropy is, given the measured kinematic data and our recovered density profile $\rho$? We need to find $prob(\alpha\vert\{data\}, \rho, K)$, where $K$ is our state of background knowledge and $\alpha$ is a quantification of velocity anisotropy. For example, it could be parametrised in terms of the anisotropy parameter $\beta$. However, the maximum likelihood approach within CHASSIS calculates $prob(f, \rho\vert\{data\}, \alpha=\alpha_0, K)$, where $f$ is the phase space density distribution that CHASSIS determines, along with $\rho$, and $\alpha_0$ is the value of $\alpha$ corresponding to isotropy in velocity space, within the adopted scheme of anisotropy parametrisation. In general, it is not possible to go from the calculated probability to the required form.

We can resort to an intermediate path by quantifying the probability of measuring a test statistic at least as extreme as the measured value of this statistic, given isotropy. This probability is referred to as a $p$-value. However, the $p$-value is a much maligned device, primarily because of the often neglected limitations of $p$-values and the subjectivity involved in establishing the acceptance of a hypothesis. Also, the $p$-value is a probability defined on sample space, but it is more satisfying to work with an alternative obtained by considering the full parameter space. 

We choose instead to employ the Bayesian $evidence$ $value$ or $ev$, details of which can be found in a well-written recent paper by \citet{pereira08} where a Fully Bayesian Significance Test (FBST) is advocated \citep{pereira99}. Our null hypothesis $H_0$ is that isotropy prevails in phase space. A brief synopsis of the FBST is presented in the Appendix, which in its full form requires the calculation of the Bayesian evidence value against $H_0$ ($\bar{ev}$), given by the integral of the posterior over the tangential set $T$.  Here $T$ comprises the mass density configurations $\rho$ that correspond to posterior probability in excess of the posterior corresponding to $\rho^{*}$, which in turn, is the point in $\rho$-space, that maximises the posterior, while satisfying $H_0$, i.e. the maximal $\rho$ that stems from the assumption of isotropy. Finally, $ev$ is obtained as $1 - {\bar{ev}}$. The definition of FBST is that the test rejects $H_0$ when $ev$ is small. To ease our calculations, we view the integral over $T$ in the conventional sense of treating probabilities, i.e. as the fractional number of cases for which $prob(\rho\vert\{data\}) > prob(\rho^{*})$. Here, the fraction is out of the total number $N$ of recorded mass density distributions; since, one mass density distribution is recorded for every iterative step, the fraction is calculated out of $N$, where $N$ is the total number of iterative steps in a run of CHASSIS. 

Our simplification assumes that the $N$ iterative steps cover the full parameter space. This may not be the case, though we need to remember that it is in \textit{proportion to} the volume of the scanned parameter space that the volume of $T$ is determined. In any case, the scanned range initiates with a seed (which has been established to be distant from the true configuration) and converges to the answer. We have also checked for the chain extending to multiple times the burn-in period as well as it being well-mixed. Thus, we can bestow confidence on the recorded density distributions covering a substantial part of the parameter space. We find that for RUN~I of CR08, $ev$=0.98, meaning our simpler version (over Pereira et. al's definition) of the Bayesian $ev$ calculation allows us to not reject $H_0$, i.e. not reject velocity isotropy as a valid assumption. In fact, following \citet{stern00}, we suggest that this high $ev$ suggests ``possibilistic support'' in favour of the assumption of isotropy.

\subsection{The globular cluster system}
It is also worth noting the interesting features of the GC system of NGC~4636. A \Chandra\ study of the level of association of GCs with LMXBs \citep{posson08} has shown consistency with similar early-type galaxies \citep[see][]{fabbiano06}. The specific frequency of GCs in NGC~4636 has consistently been found to be high \citep[$\sim$6--9, see discussion of][]{dirsch05}. \citet{dirsch05} also show that the radial distribution of GCs is shallower than the galaxy light within approximately 7$^{\prime}$. The slope changes to be consistent with the galaxy light outside 7$^{\prime}$ for the red GCs; this occurs at $\sim$9$^{\prime}$ for the blue population. Only 15 GCs are observed outside a projected radius of $\sim$ 7.5$^{\prime}$ \citep{schuberth06}. 

Our statistical analysis of the effects of orbital anisotropy leads us to accept our null hypothesis of velocity isotropy, indicating that the mass discrepancy at large radius is not the result of poorly handled anisotropy. The CHASSIS algorithm assumes the same distribution function in phase space, and therefore assumes that each GC feels the same dark matter distribution. \citet{schuberth06} suggested that the break in the radial distribution of GCs is inconsistent with NGC~4636 being embedded in a large dark matter halo. Figure 1 of CR08 shows the distribution of GCs with measured velocities, and shows that outside 30\,kpc, the radial velocities are not symmetrical about zero. More extensive coverage of GC velocities in the radial range 30--60\,kpc is required to thoroughly resolve this issue.

\subsection{X-ray systematics}
We can now examine the effects of the assumptions involved in the X-ray analysis, the most notable of which is the assumption of hydrostatic equilibrium. We also assess whether our choice of gas density model is appropriate.

\subsubsection{Hydrostatic equilibrium}
\label{sub:he}
\begin{figure}
\includegraphics[width=8cm]{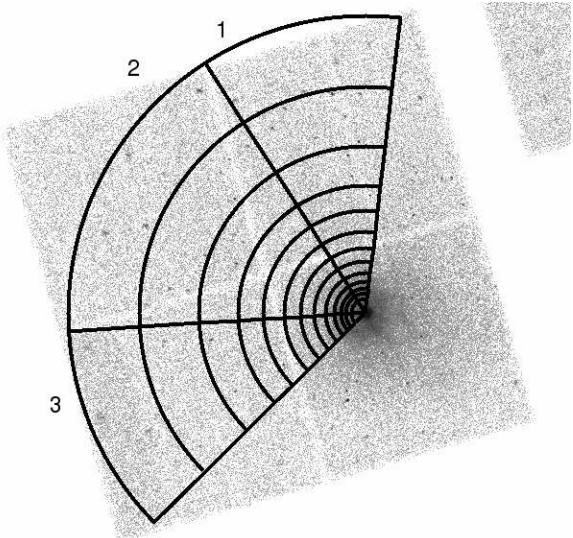}
\caption{The three regions used for extracting spectra to compare the mass profile recovered in different directions from the galaxy centre. The regions are overlaid on the filtered ACIS-I events file, binned by 4$\times$4. See Section \ref{sub:he} for details.}
\label{fig:panda}
\end{figure}
The assumption of hydrostatic equilibrium is a pre-requisite in determining the mass profile from an X-ray analysis in the manner described by Equation \ref{eqn:mass}. There is currently some controversy in this area, as recent work by \citet{diehl06} has proposed that the majority of early-type galaxy systems are not in hydrostatic equilibrium, which inevitably impacts the recovered masses. However, the results of work by \citet{churazov08} demonstrates that in mildly disturbed systems, agreement can be sought between X-ray and dynamically derived mass profiles, indicating that the assumption of hydrostatic equilibrium is indeed valid. By choosing to examine NGC~4636, we can assess the impact of any possible departures from hydrostatic equilibrium in detail. 

To assess the assumption of hydrostatic equilibrium, we have compared the recovered mass profile in three `slices' through the coarse annuli in different directions (see Figure \ref{fig:panda}). We determine the mass profiles from these regions using the \XSPEC\ \textsc{projct} model, and it is envisaged that a disturbance in the gas affecting one of these slices will be visible in the recovered mass profile in comparison to the original analysis. This implictly assumes spherical symmetry, but will give an indication of the extent to which the mass profile is affected by looking at more localised regions, instead of averaging over a full annulus.

\begin{figure*}
\begin{center}
\includegraphics[width=15cm]{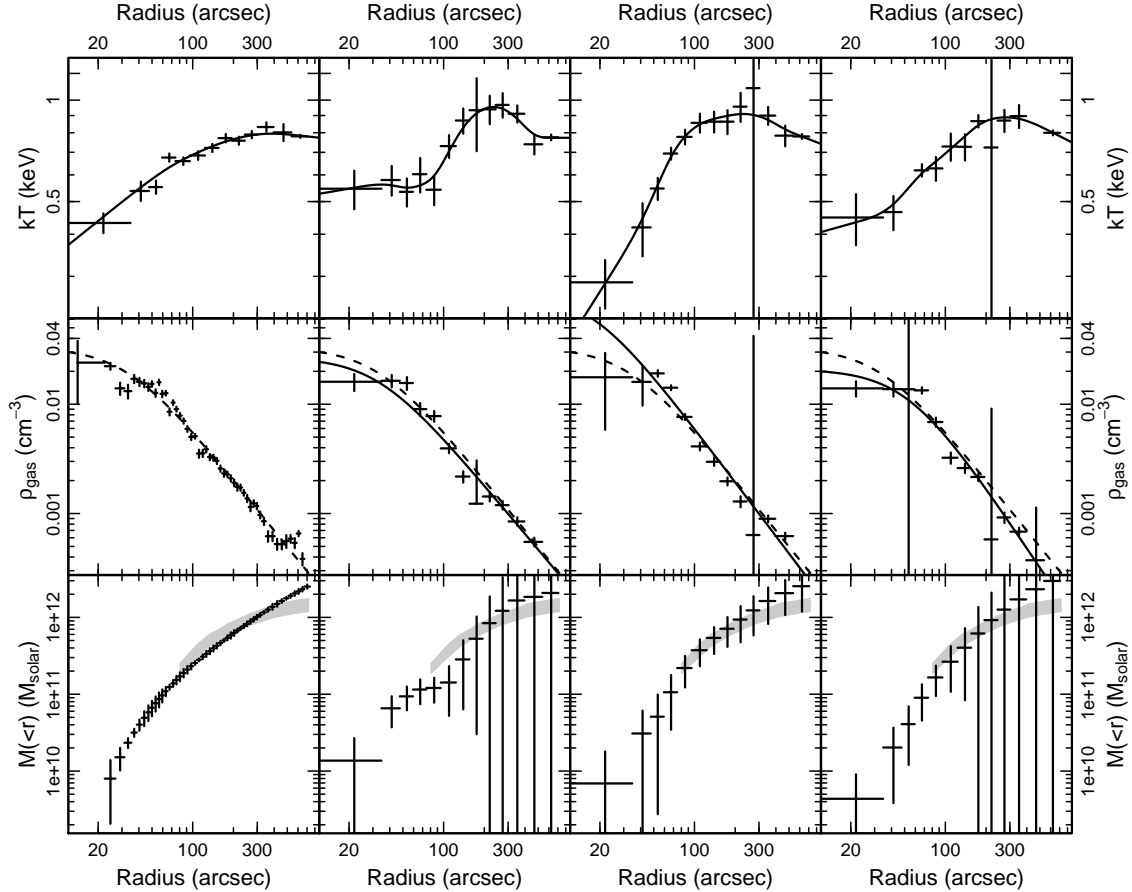}
\caption{The results of the mass analysis using the three regions described in Figure \ref{fig:panda}. The leftmost column corresponds to the original analysis and shows the deprojected temperature profile (top), gas density profile (centre) and cumulative mass profile (bottom). The remaining three columns show the deprojected temperature profiles (top), gas density profiles (centre) and cumulative mass profiles (bottom) from regions 1, 2 and 3 (Figure \ref{fig:panda}) respectively. In each deprojected temperature profile, the solid line shows the smoothing spline fit to the profile. The solid lines in each gas density profile panel shown the $\beta$-model fit to the data in that region. The dashed lines in the gas density profile panels show the same fit to the original data. The cumulative mass profiles are shown evaluated at the coarse radii. All errors shown are 1\,$\sigma$ and come from 100 MC realisations (200 in the original analysis) of the procedure. The grey confidence region in the cumulative mass profile plots shows the cumulative mass profile recovered by the dynamical analysis of the GCs by \citet{chakrabarty08}.}
\label{fig:quad}
\end{center}
\end{figure*}

Figure \ref{fig:quad} shows the recovered deprojected temperature profile, gas density profile and cumulative mass profile yielded from this analysis. The deprojection procedure in each case was more unstable than in our original procedure, due to the reduced number of counts in each spectrum, and we fixed the abundance to fit at a single value across all radii. The instability in the deprojection appears strongest in Region 3, where the penultimate temperature point is fitted very low, which if left in the fitting procedure, produces an unphysical decrease in the cumulative mass profile. We have therefore ignored this point in our smoothing spline fit. Figure \ref{fig:quad} shows a broad consistency in the inner regions, although the results from the 3 regions do differ, suggesting that as expected, the central disturbances are affecting the mass profile under the assumption of hydrostatic equilibrium.

However outside 300$^{\prime\prime}$, the recovered profiles are all consistent with the original analysis, and do not agree with the lower mass from the dynamical estimate (shown as a grey confidence region). This seems to indicate that the assumption of hydrostatic equilibrium is valid in the outskirts of this system. If localised disturbances in the gas were causing the assumption of hydrostatic equlibrium to dramatically under or over-estimate the mass this would be visible in these profiles. This is a key result; the local structure at large radius does lead to some small differences between the profiles, but the mass discrepancy between the X-ray and dynamical mass profiles \textit{is not} the result of these structural differences.

We can also examine the behaviour at small radius, where the X-ray mass is lower than the dynamically inferred mass. If we make the assumption that the dynamically inferred mass is indeed the true mass, then we can postulate what the X-ray inferred mass is telling us about bulk motions in the gas. \citet{ciotti04} show that if an X-ray analysis is applied to a situation where the gas is not in hydrostatic equlibrium, the recovered X-ray mass $M_{est}$ relates to the true mass $M$ in the following way,
\begin{equation}
M_{est} = M + \frac{r^2}{G}\nu
\end{equation}
where $G$ is the gravitational constant, $r$ is the radius and $\nu$ describes the contribution from non-hydrostatic processes. This ignores the effects of pressure terms such as those from magnetic processes. Thus, in the inner regions of the profile where the dynamically inferred mass \textit{exceeds} the X-ray mass profile, $\nu$ must be negative; in the outer regions, $\nu$ must be positive. In such a case, \citet{ciotti04} explain that in the central region, gas must be inflowing, whereas the outer region must be outflowing. The next question is whether this situation can be physically maintained. It does appear so --- \citet{pellegrini98} show that these so-called \textit{partial winds} can exist, but it seems very unlikely that this system could host an outflow with enough velocity to affect the mass estimation. Mapping the gas properties in detail outside $> 30$\,\kpc\ where we are reaching the group regime would help to settle this issue. 

Is it possible that the central region is hosting an inflow? If this is the case, it must operate on small radial scales, as the X-ray and dynamical mass profiles are consistent between $\sim$ 150$^{\prime\prime}$ and 400$^{\prime\prime}$ indicating that hydrostatic equilibrium here is obeyed. However, this is unlikely to be a long-term inflow, due to the recent (3$\times$10$^{6}$~yr) outburst from the central AGN \citep{jones02}, and limits which can be placed on the total cool gas mass \citep{sage07} suggest that the central region is not cooling to form large deposits of cool gas. It is possible that the effect of the shocks was to push gas outwards, and it is now falling back. If the presence of the shocks were affecting our spectral fits, they would raise the recovered temperature and density, and therefore would raise the mass, so the effect of excluding the shocks would be to lower the mass in the inner regions, thus worsening the discrepancy. We note that our X-ray solution which allows for the observed abundance gradient in fact reduces the recovered mass in the inner regions, due to the flattening of the gas density profile, hence worsening the agreement with the GC data in the inner regions (e.g. Figure \ref{fig:compare}).

It is also possible that the mis-match in the inner regions is the result of additional non-thermal pressure support in the gas, which could manifest from a variety of sources, such as rotation, the presence of magnetic fields, cosmic rays or a mixing of the gas with radio plasma. The presence of magnetic fields, cosmic rays or the mixing of the gas with radio plasma are all connected to the presence of a central AGN. Although the galaxy hosts a central radio source, it only extends over the central 3~kpc \citep[at 1.4\,GHz;][]{jetha07}, but there is evidence of a previous, recent AGN outburst \citep{jones02,ohto03}. Comparing the gravitational potential from the two methods can indicate the required non-thermal pressure contribution \citep{churazov08}, but the radial coverage of the GC data prevents a comparison over the radius of interest.

\subsubsection{Gas density model}
\label{sec:flat}
\begin{figure}
\includegraphics[width=8cm]{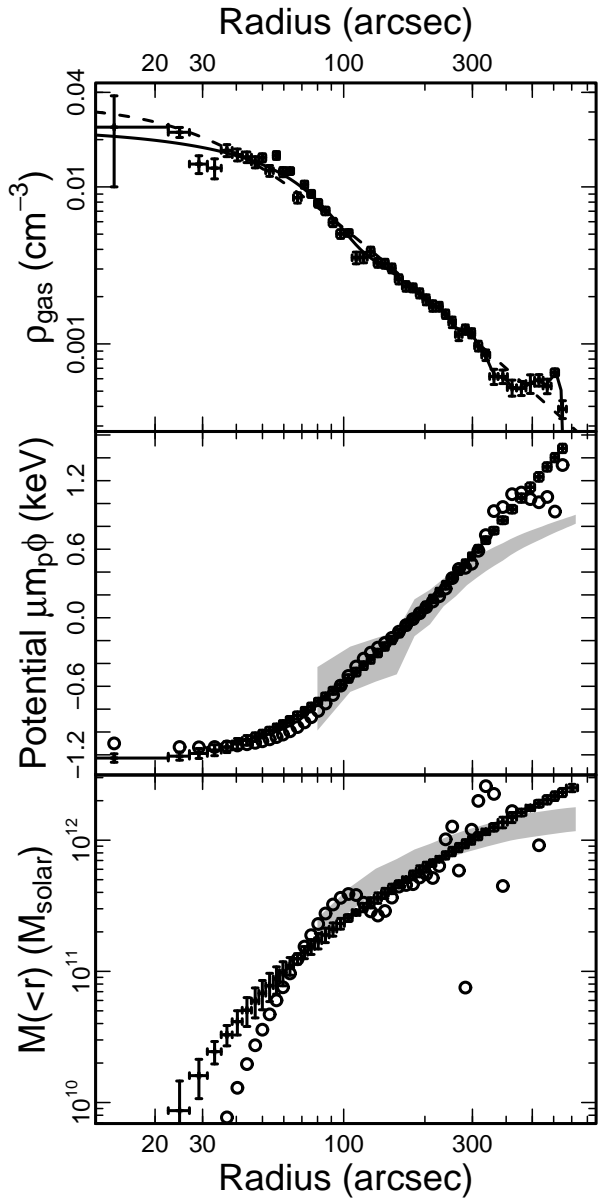}
\caption{\textit{Top panel:} The fine binned gas density profile (shown as error bars) from Figure \ref{fig:density}, with the associated $\beta$-model fit (dashed line). The smoothing spline fit to the profile (in linear--log) space is shown as the solid line. \textit{Centre panel:} The gravitational potential (in keV) recovered from the original analysis (shown as error bars, from Figure \ref{fig:pot}). The grey confidence region shows the dynamically derived potential and 1$\sigma$ errors, again from Figure \ref{fig:pot}. Open circles show the potential recovered from the smoothing spline fit to the gas density profile, shown in the top panel. \textit{Bottom panel:} The cumulative mass profile from the original analysis (shown as error bars), reproduced from Figure \ref{fig:mass}. The grey confidence region shows the dynamically recovered mass profile of CR08 (including 1$\sigma$ errors). The open circles show the mass profile resulting from the use of the smoothing spline fit to the gas density profile (shown in the top panel).}
\label{fig:flat}
\end{figure}
Figure \ref{fig:density} shows that although a $\beta$-model parameterisation of the gas density profile performs well at $\sim$100$^{\prime\prime}$--400$^{\prime\prime}$, the central regions and the outer regions are not well-described in this way. As the largest disagreement between the X-ray and dynamical profiles occurs at the largest radii, it is prudent to assess the implications of our model choice. We fit the gas density profile with a smoothing spline in linear--log space using the \rproj\ algorithm \sspline\ to better represent its shape and to allow for the local features in the profile to be incoporated into the analysis. The fit is shown in the top panel of Figure \ref{fig:flat} (solid line), alongside the original $\beta$-model (dashed line). The smoothing spline does not capture all the local features, but it performs well in representing the large radius behaviour. 

In Section \ref{sec:fine}, we noted the apparent flattening at large radius in the gas density profile reported by \citet{trinchieri94}. The difficulty in making a conclusive statement regarding the presence or not of a bump in our gas density profile lies in the inherent instability of the \textsc{projct} model in such regions, where the surface brightness profile is at its flattest \citep[see, for example][]{russell08}. This is a generic problem for deprojection schemes. In a physical system, $M(<r)$ monotonically increases, but we can see from Figure \ref{fig:flat} that allowing for a flattening in the gas density profile yields an unphysical mass profile. This is showing the limits of our deprojection, and to understand the gas properties at large radius requires mapping the gas properties to larger radius, beyond the scope of the current paper.

\section{Conclusions}
We present an X-ray mass analysis of the early-type galaxy NGC~4636 using \Chandra\ data, under the assumptions of spherical symmetry and hydrostatic equilibrium. The integrity of the latter assumption has been questioned with reference to early-type galaxies \citep{diehl06}, and it is because of the observed disturbances in the gas in NGC~4636 \citep[e.g.][]{jones02,ohto03,osullivan05b} that we chose to study this object, in an effort to assess the impact of this assumption on the recovered mass profile. We find that the treatment of the abundance gradient in the X-ray analysis can significantly affect the recovered mass profile at all radii. 

We have compared the X-ray mass density profile with that recovered from a dynamical analysis of the system's globular clusters (GCs), presented by \citet{chakrabarty08}. Inside 10\,\kpc, the dynamical mass estimate exceeds the X-ray mass estimate. The gas in this region is highly disturbed, and we postulate the cause of the disagreement to be a localised inflow of gas, or a contribution of non-thermal pressure support. 

The mass density profiles over the range $\sim$10--30\,\kpc\ are consistent within 1\,$\sigma$, indicating that even in this highly disturbed system, the recovered X-ray mass is consonant with that recovered from an independent method over intermediate radii. However, outside 30\,\kpc, the X-ray mass estimate exceeds the dynamical mass estimate, by a factor of 4--5 times at its greatest disagreement. Examining the anisotropy of the GCs, we find no statistical reason to reject our assumption of isotropy. The GC analysis is model-independent, so is not limited by the method, but the paucity of measured GC kinematics outside 7.5$^{\prime}$ means that the success of this method at large radius is limited by the data. 

We test the assumption of hydrostatic equilibrium in our X-ray analysis, finding that local disturbances at large radii do not account for the observed discrepancy. At this radius, the group gas contribution is important in this system \citep{osullivan05b}, and the overall state of the gas at this radius is uncertain. Mapping the X-ray properties to a larger radius using \XMM\ would help to model the group emission, but is beyond the scope of the current paper. 

The X-ray and dynamical mass analysis methods both indicate the need for a dark matter halo in this system, and provide a useful comparison within 30\,\kpc. It is through the comparison of independent approaches that the most robust constraints will be placed on the mass distribution of early-type galaxies, but we conclude that the limiting factors in such a comparison to large radius (outside 30\,\kpc) are data quality in the case of the GC kinematics, knowledge of the overall state of the gas as we reach the group regime in the case of the X-ray analysis, or a combination of the two.
\section*{Acknowledgments}
We are very grateful to the Royal Society for the kind contribution made towards the publication of this paper. We thank Trevor Ponman for his support and suggestions (including many fruitful discussions) throughout this project, and also for providing very helpful comments on the original manuscript. We also thank Bill Forman for useful discussions during the project. RJ thanks Alastair Sanderson for helpful discussions and for providing the software for the initial stages of the processing pipeline, and thanks Aaron Romanowsky for useful comments on the original manuscript. RJ acknowledges support from STFC/PPARC and the University of Birmingham. DC acknowledges the support of a Royal Society Dorothy Hodgkin Research Fellowship whilst at the University of Nottingham. EOS acknowledges support from NASA grants AR4-5012X and NNX07AR91G. This research has made use of the NASA/IPAC Extragalactic Database (NED) which is operated by the Jet Propulsion Laboratory, California Institute of Technology, under contract with the National Aeronautics and Space Administration.

\appendix
\section[]{Fully Bayesian Significance Test}
\noindent
The Fully Bayesian Significance Test \citep{pereira99,pereira08} requires the computation of the Bayesian evidence value ($ev$), which we define here. Let $prob(\rho)$ be the probability density function (pdf) over space $\varrho$. Let the posterior probability of $\rho$ given a measurement (represented by the data set $\{data\}$) is $prob(\rho\vert\{data\})$. Let $\rho^{*}$ be the point that maximises the posterior, while satisfying the null hypothesis $H_0$. Then, the evidence value against $H_0$ is:
\begin{equation}
{\bar{ev}} = prob(\rho \in T\vert \{data\}) ,
\end{equation}
where $T$ is the tangential set, defined as:
\begin{equation}
T = \{\rho \in \varrho : prob(\rho\vert\{data\}) > prob(\rho^{*})\}.
\end{equation}
Then ${\bar{ev}}$ can be written as:
\begin{equation}
{\bar{ev}} = \displaystyle{\int_T prob(\rho\vert\{data\})d\rho}
\end{equation}

\bibliographystyle{apj}
\bibliography{/data/ria/latex/ria_bibtex.bib}

\begin{thebibliography}{78}
\expandafter\ifx\csname natexlab\endcsname\relax\def\natexlab#1{#1}\fi

\bibitem[{{Bell} {et~al.}(2003){Bell}, {McIntosh}, {Katz}, \&
  {Weinberg}}]{bell03}
{Bell}, E.~F., {McIntosh}, D.~H., {Katz}, N., \& {Weinberg}, M.~D. 2003, ApJS,
  149, 289

\bibitem[{{Bergond} {et~al.}(2006){Bergond}, {Zepf}, {Romanowsky}, {Sharples},
  \& {Rhode}}]{bergond06}
{Bergond}, G., {Zepf}, S.~E., {Romanowsky}, A.~J., {Sharples}, R.~M., \&
  {Rhode}, K.~L. 2006, A\&A, 448, 155

\bibitem[{{Bridges} {et~al.}(2006){Bridges}, {Gebhardt}, {Sharples}, {Faifer},
  {Forte}, {Beasley}, {Zepf}, {Forbes}, {Hanes}, \& {Pierce}}]{bridges06}
{Bridges}, T., {Gebhardt}, K., {Sharples}, R., {Faifer}, F.~R., {Forte}, J.~C.,
  {Beasley}, M.~A., {Zepf}, S.~E., {Forbes}, D.~A., {Hanes}, D.~A., \&
  {Pierce}, M. 2006, MNRAS, 373, 157

\bibitem[{{Brough} {et~al.}(2006){Brough}, {Forbes}, {Kilborn}, \&
  {Couch}}]{brough06b}
{Brough}, S., {Forbes}, D.~A., {Kilborn}, V.~A., \& {Couch}, W. 2006, MNRAS,
  370, 1223

\bibitem[{Buote(2000)}]{buote00a}
Buote, D.~A. 2000, MNRAS, 311, 176

\bibitem[{{Buote} {et~al.}(2007){Buote}, {Gastaldello}, {Humphrey},
  {Zappacosta}, {Bullock}, {Brighenti}, \& {Mathews}}]{buote07}
{Buote}, D.~A., {Gastaldello}, F., {Humphrey}, P.~J., {Zappacosta}, L.,
  {Bullock}, J.~S., {Brighenti}, F., \& {Mathews}, W.~G. 2007, ApJ, 664, 123

\bibitem[{{Chakrabarty}(2006)}]{chakrabarty06}
{Chakrabarty}, D. 2006, AJ, 131, 2561

\bibitem[{{Chakrabarty} \& {Portegies Zwart}(2004)}]{chakrabarty05}
{Chakrabarty}, D. \& {Portegies Zwart}, S. 2004, AJ, 128, 1046

\bibitem[{{Chakrabarty} \& {Raychaudhury}(2008)}]{chakrabarty08}
{Chakrabarty}, D. \& {Raychaudhury}, S. 2008, AJ, 135, 2350

\bibitem[{{Chakrabarty} \& {Saha}(2001)}]{chakrabarty01}
{Chakrabarty}, D. \& {Saha}, P. 2001, AJ, 122, 232

\bibitem[{{Churazov} {et~al.}(2008){Churazov}, {Forman}, {Vikhlinin},
  {Tremaine}, {Gerhard}, \& {Jones}}]{churazov08}
{Churazov}, E., {Forman}, W., {Vikhlinin}, A., {Tremaine}, S., {Gerhard}, O.,
  \& {Jones}, C. 2008, MNRAS, 388, 1062

\bibitem[{{Ciotti} \& {Pellegrini}(2004)}]{ciotti04}
{Ciotti}, L. \& {Pellegrini}, S. 2004, MNRAS, 350, 609

\bibitem[{{Cole} {et~al.}(2001)}]{cole01}
{Cole}, S. {et~al.} 2001, MNRAS, 326, 255

\bibitem[{{C{\^o}t{\'e}} {et~al.}(2003){C{\^o}t{\'e}}, {McLaughlin}, {Cohen},
  \& {Blakeslee}}]{cote03}
{C{\^o}t{\'e}}, P., {McLaughlin}, D.~E., {Cohen}, J.~G., \& {Blakeslee}, J.~P.
  2003, ApJ, 591, 850

\bibitem[{{de Vaucouleurs} {et~al.}(1991){de Vaucouleurs}, {de Vaucouleurs},
  {Corwin}, {Buta}, {Paturel}, \& {Fouque}}]{RC3}
{de Vaucouleurs}, G., {de Vaucouleurs}, A., {Corwin}, Jr., H.~G., {Buta},
  R.~J., {Paturel}, G., \& {Fouque}, P. 1991, {Third Reference Catalogue of
  Bright Galaxies} (Volume 1-3, XII, 2069, Springer-Verlag Berlin Heidelberg
  New York)

\bibitem[{{Dekel} {et~al.}(2005){Dekel}, {Stoehr}, {Mamon}, {Cox}, {Novak}, \&
  {Primack}}]{dekel05}
{Dekel}, A., {Stoehr}, F., {Mamon}, G.~A., {Cox}, T.~J., {Novak}, G.~S., \&
  {Primack}, J.~R. 2005, Nature, 437, 707

\bibitem[{{Dickey} \& {Lockman}(1990)}]{dickey90}
{Dickey}, J.~M. \& {Lockman}, F.~J. 1990, ARA\&A, 28, 215

\bibitem[{{Diehl} \& {Statler}(2007)}]{diehl06}
{Diehl}, S. \& {Statler}, T.~S. 2007, ApJ, 668, 150

\bibitem[{{Dirsch} {et~al.}(2005){Dirsch}, {Schuberth}, \&
  {Richtler}}]{dirsch05}
{Dirsch}, B., {Schuberth}, Y., \& {Richtler}, T. 2005, A\&A, 433, 43

\bibitem[{{Douglas} {et~al.}(2007){Douglas}, {Napolitano}, {Romanowsky},
  {Coccato}, {Kuijken}, {Merrifield}, {Arnaboldi}, {Gerhard}, {Freeman},
  {Merrett}, {Noordermeer}, \& {Capaccioli}}]{douglas07}
{Douglas}, N.~G., {Napolitano}, N.~R., {Romanowsky}, A.~J., {Coccato}, L.,
  {Kuijken}, K., {Merrifield}, M.~R., {Arnaboldi}, M., {Gerhard}, O.,
  {Freeman}, K.~C., {Merrett}, H.~R., {Noordermeer}, E., \& {Capaccioli}, M.
  2007, ApJ, 664, 257

\bibitem[{{Fabbiano}(2006)}]{fabbiano06}
{Fabbiano}, G. 2006, ARA\&A, 44, 323

\bibitem[{{Fabricant} {et~al.}(1980){Fabricant}, {Lecar}, \&
  {Gorenstein}}]{fabricant80}
{Fabricant}, D., {Lecar}, M., \& {Gorenstein}, P. 1980, ApJ, 241, 552

\bibitem[{{Ferreras} {et~al.}(2008){Ferreras}, {Saha}, \&
  {Burles}}]{ferreras08}
{Ferreras}, I., {Saha}, P., \& {Burles}, S. 2008, MNRAS, 383, 857

\bibitem[{{Finoguenov} {et~al.}(2006){Finoguenov}, {Davis}, {Zimer}, \&
  {Mulchaey}}]{finoguenov06}
{Finoguenov}, A., {Davis}, D.~S., {Zimer}, M., \& {Mulchaey}, J.~S. 2006, ApJ,
  646, 143

\bibitem[{{Forman} {et~al.}(1985){Forman}, {Jones}, \& {Tucker}}]{forman85}
{Forman}, W., {Jones}, C., \& {Tucker}, W. 1985, ApJ, 293, 102

\bibitem[{{Fukazawa} {et~al.}(2006){Fukazawa}, {Botoya-Nonesa}, {Pu}, {Ohto},
  \& {Kawano}}]{fukazawa06}
{Fukazawa}, Y., {Botoya-Nonesa}, J.~G., {Pu}, J., {Ohto}, A., \& {Kawano}, N.
  2006, ApJ, 636, 698

\bibitem[{{Gebhardt} {et~al.}(2000){Gebhardt}, {Pryor}, {O'Connell},
  {Williams}, \& {Hesser}}]{gebhardt00}
{Gebhardt}, K., {Pryor}, C., {O'Connell}, R.~D., {Williams}, T.~B., \&
  {Hesser}, J.~E. 2000, AJ, 119, 1268

\bibitem[{{Genzel} {et~al.}(2000){Genzel}, {Pichon}, {Eckart}, {Gerhard}, \&
  {Ott}}]{genzel00}
{Genzel}, R., {Pichon}, C., {Eckart}, A., {Gerhard}, O.~E., \& {Ott}, T. 2000,
  MNRAS, 317, 348

\bibitem[{{Ghez} {et~al.}(1998){Ghez}, {Klein}, {Morris}, \&
  {Becklin}}]{ghez98}
{Ghez}, A.~M., {Klein}, B.~L., {Morris}, M., \& {Becklin}, E.~E. 1998, ApJ,
  509, 678

\bibitem[{{Grevesse} \& {Sauval}(1998)}]{grevesse98}
{Grevesse}, N. \& {Sauval}, A.~J. 1998, Space Science Reviews, 85, 161

\bibitem[{{Hoekstra} {et~al.}(2005){Hoekstra}, {Hsieh}, {Yee}, {Lin}, \&
  {Gladders}}]{hoekstra05}
{Hoekstra}, H., {Hsieh}, B.~C., {Yee}, H.~K.~C., {Lin}, H., \& {Gladders},
  M.~D. 2005, ApJ, 635, 73

\bibitem[{{Humphrey} {et~al.}(2006){Humphrey}, {Buote}, {Gastaldello},
  {Zappacosta}, {Bullock}, {Brighenti}, \& {Mathews}}]{humphrey06}
{Humphrey}, P.~J., {Buote}, D.~A., {Gastaldello}, F., {Zappacosta}, L.,
  {Bullock}, J.~S., {Brighenti}, F., \& {Mathews}, W.~G. 2006, ApJ, 646, 899

\bibitem[{{Irwin} {et~al.}(2003){Irwin}, {Athey}, \& {Bregman}}]{irwin03}
{Irwin}, J.~A., {Athey}, A.~E., \& {Bregman}, J.~N. 2003, ApJ, 587, 356

\bibitem[{{Jarrett} {et~al.}(2003){Jarrett}, {Chester}, {Cutri}, {Schneider},
  \& {Huchra}}]{jarrett03}
{Jarrett}, T.~H., {Chester}, T., {Cutri}, R., {Schneider}, S.~E., \& {Huchra},
  J.~P. 2003, AJ, 125, 525

\bibitem[{{Jetha} {et~al.}(2007){Jetha}, {Ponman}, {Hardcastle}, \&
  {Croston}}]{jetha07}
{Jetha}, N.~N., {Ponman}, T.~J., {Hardcastle}, M.~J., \& {Croston}, J.~H. 2007,
  MNRAS, 376, 193

\bibitem[{{Jones} {et~al.}(2002){Jones}, {Forman}, {Vikhlinin}, {Markevitch},
  {David}, {Warmflash}, {Murray}, \& {Nulsen}}]{jones02}
{Jones}, C., {Forman}, W., {Vikhlinin}, A., {Markevitch}, M., {David}, L.,
  {Warmflash}, A., {Murray}, S., \& {Nulsen}, P.~E.~J. 2002, ApJ, 567, L115

\bibitem[{{Kim} \& {Fabbiano}(2004)}]{kim04}
{Kim}, D.-W. \& {Fabbiano}, G. 2004, ApJ, 611, 846

\bibitem[{{Kim} {et~al.}(2006){Kim}, {Kim}, {Fabbiano}, {Lee}, {Park},
  {Geisler}, \& {Dirsch}}]{kim06}
{Kim}, E., {Kim}, D.-W., {Fabbiano}, G., {Lee}, M.~G., {Park}, H.~S.,
  {Geisler}, D., \& {Dirsch}, B. 2006, ApJ, 647, 276

\bibitem[{{Kleinheinrich} {et~al.}(2006){Kleinheinrich}, {Schneider}, {Rix},
  {Erben}, {Wolf}, {Schirmer}, {Meisenheimer}, {Borch}, {Dye}, {Kovacs}, \&
  {Wisotzki}}]{kleinheinrich06}
{Kleinheinrich}, M., {Schneider}, P., {Rix}, H.-W., {Erben}, T., {Wolf}, C.,
  {Schirmer}, M., {Meisenheimer}, K., {Borch}, A., {Dye}, S., {Kovacs}, Z., \&
  {Wisotzki}, L. 2006, A\&A, 455, 441

\bibitem[{{Koopmans}(2006)}]{koopmans06}
{Koopmans}, L.~V.~E. 2006, in EAS Publications Series, Vol.~20, EAS
  Publications Series, ed. G.~A. {Mamon}, F.~{Combes}, C.~{Deffayet}, \&
  B.~{Fort}, 161--166

\bibitem[{{Loewenstein} \& {Mushotzky}(2003)}]{loewenstein03}
{Loewenstein}, M. \& {Mushotzky}, R. 2003, Nucl. Phys. B (Proc. Suppl.), 124,
  91

\bibitem[{{{\L}okas} {et~al.}(2007){{\L}okas}, {Wojtak}, {Mamon}, \&
  {Gottloeber}}]{lokas07}
{{\L}okas}, E.~L., {Wojtak}, R., {Mamon}, G.~A., \& {Gottloeber}, S. 2007,
  arXiv: 0712.2368

\bibitem[{{Magorrian} \& {Ballantyne}(2001)}]{magorrian01}
{Magorrian}, J. \& {Ballantyne}, D. 2001, MNRAS, 322, 702

\bibitem[{{Mamon} \& {{\L}okas}(2005)}]{mamon05a}
{Mamon}, G.~A. \& {{\L}okas}, E.~L. 2005, MNRAS, 362, 95

\bibitem[{{Mandelbaum} {et~al.}(2006){Mandelbaum}, {Hirata}, {Broderick},
  {Seljak}, \& {Brinkmann}}]{mandelbaum06}
{Mandelbaum}, R., {Hirata}, C.~M., {Broderick}, T., {Seljak}, U., \&
  {Brinkmann}, J. 2006, MNRAS, 370, 1008

\bibitem[{{McLaughlin}(1999)}]{mclaughlin99}
{McLaughlin}, D.~E. 1999, AJ, 117, 2398

\bibitem[{Navarro {et~al.}(1996)Navarro, Frenk, \& White}]{navarro96}
Navarro, J.~F., Frenk, C.~S., \& White, S.~D.~M. 1996, ApJ, 462, 563

\bibitem[{{Nolthenius}(1993)}]{nolthenius93}
{Nolthenius}, R. 1993, ApJS, 85, 1

\bibitem[{{Ohto} {et~al.}(2003){Ohto}, {Kawano}, \& {Fukazawa}}]{ohto03}
{Ohto}, A., {Kawano}, N., \& {Fukazawa}, Y. 2003, PASJ, 55, 819

\bibitem[{{Osmond} \& {Ponman}(2004)}]{osmond04}
{Osmond}, J.~P.~F. \& {Ponman}, T.~J. 2004, MNRAS, 350, 1511

\bibitem[{O'Sullivan {et~al.}(2001)O'Sullivan, Forbes, \&
  Ponman}]{osullivan01b}
O'Sullivan, E., Forbes, D.~A., \& Ponman, T.~J. 2001, MNRAS, 328, 461

\bibitem[{{O'Sullivan} \& {Ponman}(2004)}]{osullivan04b}
{O'Sullivan}, E. \& {Ponman}, T.~J. 2004, MNRAS, 354, 935

\bibitem[{{O'Sullivan} {et~al.}(2005){O'Sullivan}, {Vrtilek}, \&
  {Kempner}}]{osullivan05b}
{O'Sullivan}, E., {Vrtilek}, J.~M., \& {Kempner}, J.~C. 2005, ApJ, 624, L77

\bibitem[{{Paturel} {et~al.}(2003){Paturel}, {Petit}, {Prugniel}, {Theureau},
  {Rousseau}, {Brouty}, {Dubois}, \& {Cambr{\'e}sy}}]{paturel03}
{Paturel}, G., {Petit}, C., {Prugniel}, P., {Theureau}, G., {Rousseau}, J.,
  {Brouty}, M., {Dubois}, P., \& {Cambr{\'e}sy}, L. 2003, A\&A, 412, 45

\bibitem[{Pellegrini \& Ciotti(1998)}]{pellegrini98}
Pellegrini, S. \& Ciotti, L. 1998, A\&A, 333, 433

\bibitem[{{Pellegrini} \& {Ciotti}(2006)}]{pellegrini06}
{Pellegrini}, S. \& {Ciotti}, L. 2006, MNRAS, 370, 1797

\bibitem[{{Pereira} \& {Stern}(1999)}]{pereira99}
{Pereira}, C. \& {Stern}, J. 1999, {Entropy}, 1, 99

\bibitem[{{Pereira} {et~al.}(2008){Pereira}, {Stern}, \&
  {Wechsler}}]{pereira08}
{Pereira}, C., {Stern}, J., \& {Wechsler}, S. 2008, {Bayesian Analysis}, 3, 79

\bibitem[{{Piffaretti} {et~al.}(2003){Piffaretti}, {Jetzer}, \&
  {Schindler}}]{piffaretti03}
{Piffaretti}, R., {Jetzer}, P., \& {Schindler}, S. 2003, A\&A, 398, 41

\bibitem[{{Posson-Brown} {et~al.}(2009){Posson-Brown}, {Raychaudhury},
  {Forman}, Donnelly, \& {Jones}}]{posson08}
{Posson-Brown}, J., {Raychaudhury}, S., {Forman}, W., Donnelly, R., \& {Jones},
  C. 2009, ApJ, accepted, astro-ph/0605308

\bibitem[{{Prugniel} \& {Simien}(1997)}]{prugniel97}
{Prugniel}, P. \& {Simien}, F. 1997, A\&A, 321, 111

\bibitem[{{R Development Core Team}(2008)}]{Rcite}
{R Development Core Team}. 2008, R: A Language and Environment for Statistical
  Computing, R Foundation for Statistical Computing, Vienna, Austria, {ISBN}
  3-900051-07-0

\bibitem[{{Rasmussen} \& {Ponman}(2007)}]{rasmussen07}
{Rasmussen}, J. \& {Ponman}, T.~J. 2007, MNRAS, 380, 1554

\bibitem[{{Romanowsky} {et~al.}(2003){Romanowsky}, {Douglas}, {Arnaboldi},
  {Kuijken}, {Merrifield}, {Napolitano}, {Capaccioli}, \&
  {Freeman}}]{romanowsky03}
{Romanowsky}, A.~J., {Douglas}, N.~G., {Arnaboldi}, M., {Kuijken}, K.,
  {Merrifield}, M.~R., {Napolitano}, N.~R., {Capaccioli}, M., \& {Freeman},
  K.~C. 2003, Science, 301, 1696

\bibitem[{{Romanowsky} \& {Kochanek}(2001)}]{romanowsky01}
{Romanowsky}, A.~J. \& {Kochanek}, C.~S. 2001, ApJ, 553, 722

\bibitem[{{Romanowsky} {et~al.}(2009){Romanowsky}, {Strader}, {Spitler},
  {Johnson}, {Brodie}, {Forbes}, \& {Ponman}}]{romanowsky08}
{Romanowsky}, A.~J., {Strader}, J., {Spitler}, L.~R., {Johnson}, R., {Brodie},
  J.~P., {Forbes}, D.~A., \& {Ponman}, T. 2009, AJ, 137, 4956

\bibitem[{{Russell} {et~al.}(2008){Russell}, {Sanders}, \&
  {Fabian}}]{russell08}
{Russell}, H.~R., {Sanders}, J.~S., \& {Fabian}, A.~C. 2008, MNRAS, 390, 1207

\bibitem[{{Sage} {et~al.}(2007){Sage}, {Welch}, \& {Young}}]{sage07}
{Sage}, L.~J., {Welch}, G.~A., \& {Young}, L.~M. 2007, ApJ, 657, 232

\bibitem[{{Sambhus} {et~al.}(2006){Sambhus}, {Gerhard}, \&
  {M{\'e}ndez}}]{sambhus06}
{Sambhus}, N., {Gerhard}, O., \& {M{\'e}ndez}, R.~H. 2006, AJ, 131, 837

\bibitem[{{Samurovi{\'c}} \& {Danziger}(2006)}]{samurovic06}
{Samurovi{\'c}}, S. \& {Danziger}, I.~J. 2006, A\&A, 458, 79

\bibitem[{{Schuberth} {et~al.}(2006){Schuberth}, {Richtler}, {Dirsch},
  {Hilker}, {Larsen}, {Kissler-Patig}, \& {Mebold}}]{schuberth06}
{Schuberth}, Y., {Richtler}, T., {Dirsch}, B., {Hilker}, M., {Larsen}, S.~S.,
  {Kissler-Patig}, M., \& {Mebold}, U. 2006, A\&A, 459, 391

\bibitem[{{Sheldon} {et~al.}(2004){Sheldon}, {Johnston}, {Frieman}, {Scranton},
  {McKay}, {Connolly}, {Budav{\'a}ri}, {Zehavi}, {Bahcall}, {Brinkmann}, \&
  {Fukugita}}]{sheldon04}
{Sheldon}, E.~S., {Johnston}, D.~E., {Frieman}, J.~A., {Scranton}, R., {McKay},
  T.~A., {Connolly}, A.~J., {Budav{\'a}ri}, T., {Zehavi}, I., {Bahcall}, N.~A.,
  {Brinkmann}, J., \& {Fukugita}, M. 2004, AJ, 127, 2544

\bibitem[{{Sofue} \& {Rubin}(2001)}]{sofue01}
{Sofue}, Y. \& {Rubin}, V. 2001, ARA\&A, 39, 137

\bibitem[{{Stern}(2000)}]{stern00}
{Stern}, J. 2000, {Frontiers in Artificial Intelligence and its Applications},
  101, 139

\bibitem[{{Trinchieri} {et~al.}(1994){Trinchieri}, {Kim}, {Fabbiano}, \&
  {Canizares}}]{trinchieri94}
{Trinchieri}, G., {Kim}, D.-W., {Fabbiano}, G., \& {Canizares}, C.~R.~C. 1994,
  ApJ, 428, 555

\bibitem[{{Vikhlinin} {et~al.}(2006){Vikhlinin}, {Kravtsov}, {Forman}, {Jones},
  {Markevitch}, {Murray}, \& {Van Speybroeck}}]{vikhlinin06}
{Vikhlinin}, A., {Kravtsov}, A., {Forman}, W., {Jones}, C., {Markevitch}, M.,
  {Murray}, S.~S., \& {Van Speybroeck}, L. 2006, ApJ, 640, 691

\bibitem[{{Woodley} {et~al.}(2007){Woodley}, {Harris}, {Beasley}, {Peng},
  {Bridges}, {Forbes}, \& {Harris}}]{woodley07}
{Woodley}, K.~A., {Harris}, W.~E., {Beasley}, M.~A., {Peng}, E.~W., {Bridges},
  T.~J., {Forbes}, D.~A., \& {Harris}, G.~L.~H. 2007, AJ, 134, 494

\bibitem[{{Zhang} {et~al.}(2007){Zhang}, {Xu}, {Wang}, {An}, {Xu}, \&
  {Wu}}]{zhang07}
{Zhang}, Z., {Xu}, H., {Wang}, Y., {An}, T., {Xu}, Y., \& {Wu}, X.-P. 2007,
  ApJ, 656, 805

\end{thebibliography}
\end{document}